\documentclass{aa}
\usepackage{graphicx}
\usepackage{txfonts} 
\input amssym.tex
\def\erv{{\bf e}_r}
\def\xv{{\bf x}}
\def\rb{r_b}
\def\rt{r_t}
\def\Rd{R_d}
\def\Eg{E_g}
\def\Egb{E_{gb}}
\def\EgDM{E_{gDM}}
\def\Egd{E_{gd}}
\def\Wb{W_b}
\def\Wbb{W_{bb}}
\def\WbDM{W_{bDM}}
\def\Wbd{W_{bd}}

\begin{document}
 
\title{Evolution of chemical abundances in Seyfert galaxies}

\author{Silvia K. Ballero\inst{1,2}, 
Francesca Matteucci\inst{1,2},
Luca Ciotti\inst{3},
Francesco Calura\inst{2}, 
Paolo Padovani\inst{4}}

\offprints{\ttfamily ballero@oats.inaf.it}

\institute{Dipartimento di Astronomia, Universit\`a di Trieste, via
  G.B. Tiepolo 11, 34143 Trieste, Italy
\and
  INAF, Osservatorio Astronomico di Trieste, via G.B. Tiepolo 11, 34143
  Trieste, Italy
\and
  Dipartimento di Astronomia, Universit\`a di Bologna, via Ranzani 1, 40127,
  Bologna, Italy
\and
  European Organisation for Astronomical Research in the Southern 
Hemisphere (ESO), Karl-Schwarzschild-Str. 2, 85748
  Garching bei M\"unchen, Germany}

\date{Draft, October 19, 2007}

\abstract{}
{To compute the chemical evolution of spiral bulges hosting Seyfert
  nuclei, based on updated chemical and spectro-photometrical
  evolution models for the bulge of our Galaxy, to make predictions
  about other quantities measured in Seyferts, and to model the
  photometric features of local bulges. 
  The chemical evolution model contains updated and detailed
  calculations of the Galactic potential and of the feedback from the
  central supermassive black hole, and the spectro-photometric model
  covers a wide range of stellar ages and metallicities.}
{We computed the evolution of bulges in the mass range $2\times
  10^{9}-10^{11}M_{\odot}$ by scaling the efficiency of star formation
  and the bulge scalelength as in the inverse-wind scenario for
  elliptical galaxies, and considering an Eddington limited accretion
  onto the central supermassive black hole.} 
{We successfully reproduced the observed relation between the mass of
  the black hole and that of the host bulge. 
  The observed nuclear bolometric luminosity emitted by the
  supermassive black hole is reproduced only at high redshift or for
  the most massive bulges; in the other cases, at $z \simeq 0$ a
  rejuvenation mechanism is necessary.
  The energy provided by the black hole is in most cases not
  significant in the triggering of the galactic wind. 
  The observed high star formation rates and metal overabundances are
  easily achieved, as well as the constancy of chemical abundances
  with the redshift and the bulge present-day colours.
  Those results are not affected if we vary the index of the stellar
  IMF from $x=0.95$ to $x=1.35$; a steeper IMF is instead required in
  order to reproduce the colour-magnitude relation and the present
  $K$-band luminosity of the bulge.}
{We show that the chemical evolution of the host bulge, with a short
  formation timescale of $\sim 0.1$ Gyr, a rather high efficiency of
  star formation ranging from $11$ to $50$ Gyr$^{-1}$ according to the 
  bulge mass and an IMF flatter with respect to the solar
  neighbourhood, combined with the accretion onto the black hole is
  sufficient to explain the main observed features of Seyfert
  galaxies.}


\titlerunning{Evolution of Seyfert galaxies}
\authorrunning{S.K. Ballero et al.}

\maketitle

\section{Introduction}

The outstanding question of the co-evolution of Active Galactic Nuclei
(AGNs) and their host galaxies has received considerable attention in
the past decades, since various pieces of evidence pointed to a link
between the formation of supermassive black holes (BHs) and the
formation and evolution of their host spheroids: for example, the
usual presence of massive dark objects at the centre of nearby
spheroids (Ford et al. 1997; Ho 1999; Wandel 1999); the correlation
between the BH mass and the stellar velocity dispersion of the host
(for quiescent galaxies, Ferrarese \& Merritt 2000; Gebhardt et
al. 2000a; Tremaine et al. 2002; for active galaxies, Gebhardt et
al. 2000b; Ferrarese et al. 2001; Shields et al., 2003; Onken et
al. 2004; Nelson et al. 2004) or its mass (Kormendy \& Richstone 1995;
Magorrian et al. 1998; Marconi \& Hunt 2003; Dunlop et al. 2003); the
similarity between light evolution of quasar (QSO) population and the
star formation history of galaxies (Cavaliere \& Vittorini 1998;
Haiman et al. 2004); the establishment of a good match
among the optical QSO luminosity function, the luminosity function of
star-forming galaxies and the mass function of dark matter halos
(DMHs) at $z\sim 3$ (Haenhelt et al., 1998).

The most widely accepted explanation for the luminosity emitted by an
AGN, is radiatively efficient gas accretion onto a central supermassive
BH.
The outflows from AGNs can profoundly affect the evolution of the host
galaxy, e.g. by quenching or inducing the star formation (e.g., see
Ciotti \& Ostriker 2007, and references therein).  The mutual feedback
between galaxies and QSOs was used as a key to solve the shortcomings
of the semianalytic models in galaxy evolution, e.g. the failure to
account for the surface density of high-redshift massive galaxies
(Blain et al., 2002; Cimatti et al., 2002) and for the
$\alpha$-enhancement as a function of mass (Thomas et al., 2002),
since it could provide a way to invert the hierarchical scenario for
the assembly of galaxies and star formation (see e.g. Monaco et al.,
2000; Granato et al., 2004; Scannapieco et al., 2005).

The study of the chemical abundances of the QSOs was first undertaken
by Hamann \& Ferland (1993), who combined chemical evolution and
spectral synthesis models to interpret the N~{\scshape v}/C~{\scshape
iv} and N~{\scshape v}/He~{\scshape ii} broad emission line ratios,
and found out that the high metallicities and the abundance ratios of
the broad-line region are consistent with the outcomes of the models
for giant elliptical galaxies (Arimoto \& Yoshii, 1987; Matteucci \&
Tornamb\`e, 1987; Angeletti \& Giannone, 1990), where the timescales
of star formation and enrichment are very short and the initial mass
function (IMF) is top-heavy.  In the same year, Padovani \& Matteucci
(1993) and Matteucci \& Padovani (1993) employed the chemical
evolution model of Matteucci (1992) to model the evolution of
radio-loud QSOs, which are hosted by massive ellipticals, following in
detail the evolution of several chemical species in the gas.  They
supposed that the mass loss from dying stars after the galactic wind
provides the fuel for the central BH and modeled the bolometric
luminosity as $L_{bol}=\eta\dot{M}c^2$, with a typical value for the
efficiency of $\eta = 0.1$ and were successful in obtaining the
estimated QSO luminosities and the observed ratio of AGN to host
galaxy luminosity.  Then, they studied the evolution of the chemical
composition of the gas lost by stars in elliptical galaxies and spiral
bulges for various elements (C, N, O, Ne, Mg, Si and Fe), and found
out that due to the high star-formation rate (SFR) of spheroids at
early times the standard QSO emission lines were naturally explained.
The relatively weak observed time dependence of the QSO abundances for
$t \gtrsim 1$ Gyr was also predicted.  The model of Matteucci \&
Padovani (1993) still followed the classic wind scenario, where the
efficiency of star formation decreases with increasing galactic mass
and which was found to be inconsistent with the correlation between
spheroid mass and $\alpha$-enhancement (Matteucci 1994).  Moreover,
Padovani \& Matteucci (1993) pointed out that if all mass lost 
by stars in the host galaxy after the wind were accreted by the
central BH, the final BH mass would be up 
to two orders of magnitude larger than observed.

Other works (Fria\c ca \& Terlevich 1998; Romano et al. 2002; Granato
et al., 2004), which had a more refined treatment of gas dynamics,
limited their analysis of chemical abundances to the metallicity $Z$
and the [Mg/Fe] ratio and their correlation with the galactic mass.

All these studies were mainly devoted to studying the co-evolution of
radio-loud QSOs and their host spheroids, which are elliptical
galaxies.  Now we want to extend the approach of Padovani \& Matteucci
(1993) to AGNs hosted by spiral bulges, with a more recent chemical
evolution model for the bulge with the introduction of the treatment
of feedback from the central BH and a more sophisticated dealing of
the accretion rate.  Since Seyfert nuclei are preferentially hosted by
disk-dominated galaxies (Adams, 1977; Yee, 1983; MacKenty, 1990; Ho et
al. 1997) our study can be applied to this class of objects.

The paper is organized as follows: in \S 2 we illustrate the chemical
and photometrical evolution model, in \S 3 we show our calculations of
the potential energy and of the feedback from supernovae (SNe) and
from the AGN, in \S 4 we discuss our results concerning the black hole
masses and luminosities, the chemical abundances and the photometry,
and in \S 5 we draw some conclusions.

\section{The evolution model}

\subsection{Chemical evolution}
\label{sec:chem}

The model on which we base our analysis is essentially the recent
bulge chemical evolution model from Ballero et al. (2007a), which was
successful in reproducing the most recent measurements of metallicity
distribution (Zoccali et al., 2003; Fulbright et al., 2006) and
evolution of abundance ratios (Origlia et al., 2002; Origlia \& Rich,
2004; Origlia et al., 2005; Rich \& Origlia, 2005; Fulbright et al.,
2007; Zoccali et al., 2007; Lecureur et al., 2007) of the bulge
giants.  Here we resume its main features.

\begin{itemize}

\item[-] The gas is supposed to be well-mixed and homogeneous at any
  time (instantaneous mixing approximation).

\item[-] The star formation rate (SFR) is parametrized as 

\begin{equation}
\psi (r,t) \propto \nu \sigma_{gas}^k(r,t).
\end{equation}

In this expression $\nu$ is the efficiency of star formation (a
parameter whose meaning is that of the inverse of the timescale of
star formation), $\sigma_{gas}$ is the gas surface mass density
and $k$ is an index whose value can vary between $1$ and $1.5$ without giving rise to remarkable differences.  We chose $k=1$ to recover the star
formation law of spheroids. 
In our formalism the SFR is in units of $M_{\odot}$Gyr$^{-1}$pc$^{-2}$.

\item[-] The bulge forms by accreting gas from the Galactic halo at an
  exponential rate:

\begin{equation} 
\dot{M}_{inf} \propto e^{-t/\tau},
\end{equation} 
where $\tau$ is a suitable collapse timescale.  The metallicity
$Z_{acc}$ of the accreted gas is of the order of $\lesssim
10^{-4}Z_{\odot}$, but it is easily shown that the results are not
significantly affected if we consider $Z_{acc} = 0$.

\item[-] The Type Ia SN rate is computed like in Matteucci \&
Recchi (2001) following the single degenerate scenario of Nomoto et
al. (1984).

\item[-] Stellar lifetimes (Kodama, 1997) are taken into account in
detail. We adopted the stellar yields of Fran\c cois et al. (2004),
which were constrained in order to reproduce the observed chemical
abundances of the solar neighbourhood in the two-infall
model of Chiappini et al. (2003).

\item[-] The adopted IMF is constant in space and time and has the
shape of a multi-part power law:
\begin{equation}
  \phi(M)\propto M^{-(1+x_i)}
\end{equation}
where the subscript refers to different mass ranges. Its normalization
is performed by assuming a stellar mass range of $0.1-80M_{\odot}$.

\end{itemize}

In the reference model of Ballero et al. (2007a), the parameters which
allow to best fit the metallicity distributions and the [$\alpha$/Fe]
\emph{vs.} [Fe/H] ratios measured in the bulge giants are the
following: $\nu=20$ Gyr$^{-1}$, $\tau=0.1$ Gyr, and two slopes for the
IMF, namely $x_1 = 0.33$ for $0.1 \leq M/M_{\odot} \leq 1$ (in
agreement with the photometric measurements of Zoccali et al., 2000)
and $x_2 = 0.95$ for $1 \leq M/M_{\odot} \leq 80$.  We also considered
the case $x_{2}=1.35$ (Salpeter, 1955) for comparison.

This model holds for a galaxy like ours with a bulge of $M_{b} =
2\times 10^{10}M_{\odot}$.  We are going to predict the properties of
Seyfert nuclei hosted by bulges of different masses, therefore some
model parameters will have to be re-scaled. We choose to keep the
IMF constant and to scale the effective
radius and the star formation efficiency following the inverse-wind
scenario (Matteucci, 1994).  
The possibility of changing the infall timescale with mass is not
explored in the present paper. 
In Table \ref{tab:tab1} are reported the adopted parameters for each
bulge mass.

\begin{table}
\centering
\footnotesize
\begin{tabular}{rccc}
\hline
\hline
$M_{b}$ ($M_{\odot}$)& 
$\nu$ (Gyr$^{-1}$) & 
$R_{e}$ (kpc) &
$t_{GW}$ (Gyr) \\
\hline
$2 \times 10^{9\phantom{0}}$ & 11 & 1 & 0.31 \\
$2 \times 10^{10}$         & 20 & 2 & 0.27 \\
         $10^{11}$         & 50 & 4 & 0.22 \\
\hline
\end{tabular}
\caption{Features of the examined models: bulge mass (first column),
  star formation efficiency (second column), bulge effective radius
  (third column). The table also reports the time of occurrence of the
  galactic wind (fourth column).}
\label{tab:tab1}
\end{table}

\subsection{The spectro-photometric model}

By matching chemical evolution models with a spectro-photometric code,
it has been possible to reproduce the present-day photometric features
of galaxies of various morphological types (Calura \& Matteucci 2006,
Calura et al. 2007a) and to perform detailed studies of
the evolution of the luminous matter in the Universe (Calura \&
Matteucci 2003, Calura et al. 2004).  In this paper, by
means of the chemical evolution model plus a spectro-photometric code,
we attempt to model the photometric features of galactic bulges.  All
the spectro-photometric calculations are performed by means of the
code developed by Jimenez et al. (2004); this model is based on new
stellar isochrones computed by Jimenez et al. (1998) and on the
stellar atmospheric models by Kurucz (1992).  The main advantage of
this photometric code is that it allows one to follow in detail the
metallicity evolution of the gas; this is possible thanks to the large
number of simple stellar populations calculated by Jimenez et
al. (2004) by means of new stellar tracks, with ages between $1 \times
10^{6}$ and $1.4 \times 10^{10}$ yr and metallicities ranging from
$Z=0.0002$ to $Z=0.1$.

Starting from the stellar spectra, first we build simple stellar
population (SSP) models consistent with the chemical evolution at any
given time and weighted according to the assumed IMF.  Then, a
composite stellar population (CSP) consists of the sum of different
SSPs formed at different times, with a luminosity at an age $t_{0}$
and at a particular wavelength $\lambda$ given by:
\begin{equation}
L_{\lambda}(t_{0})=\int_{0}^{t_{0}} \int_{Z_{i}}^{Z_{f}} 
                   \psi(t_{0}-t) L_{SSP,\lambda}(Z,t_{0}-t)dZdt
\end{equation}
where the luminosity of the SSP can be written as:
\begin{equation}
 L_{SSP,\lambda}(Z, t_{0}-t)= \int_{M_{min}}^{M_{max}} 
                   \phi(M) l_{\lambda}(Z,M,t_{0}-t)dM
\end{equation}
and where $l_{\lambda}(Z,M,t_{0}-t)$ is the luminosity of a star of
mass $M$, metallicity $Z$ and age $t_{0}-t$; $Z_{i}$ and $Z_{f}$ are
the initial and final metallicities, $M_{min}$ and $M_{max}$ are the
smallest and largest stellar mass in the population, $\phi(M)$ is the
IMF and $\psi(t)$ is the SFR at the time t.  
Peletier et al. (1999) have shown that dust in local bulges is very 
patchy and concentrated in the innermost regions, 
i.e. within distances of $\sim 100 pc$.  They have also shown that dust 
extinction effects are negligible at distances 
of $\sim 1 R_{eff}$.  
In addition, we compare our photometric predictions to observational results 
largely unaffected by dust, such as the ones by Balcells \& Peletier 1994. 
For these reasons, in all our
spectro-photometric calculations, we do not take into account dust
extinction.

\section{Energetics and AGN feedback}

The bulge lies in the potential well of the Galaxy, which consists of
both luminous and dark matter.  In the model of Ballero et al.~(2007a)
we calculated the binding energy $\Delta\Eg$ of the bulge gas
following Bertin et al. (1992), i.e. by treating the bulge as a
scaled-down two-component elliptical (thus ignoring the disk
contribution).  We also assumed that the thermal energy of the bulge
interstellar medium was mainly contributed by the explosion of Type I
and Type II supernovae. In particular, if we call $E_{th}$ the gas
thermal energy, we supposed that at the time $t_{GW}$, when the
condition
\begin{equation}
E_{th}(t_{GW})=\Delta\Eg (t_{GW})
\end{equation}
is satisfied, a wind develops and the star formation is suppressed.
In the reference model of Ballero et al. (2007a), this event does not
have a great impact on the predicted chemical evolution, since it
occurs when most of the gas has already been processed into stars.
However, the high-metallicity tail of the metallicity distribution was
somewhat overestimated without the galactic~wind.

In the present exploration not only do we adopt a more realistic
disk galaxy model to better estimate the binding energy of the gas in
the bulge, but we also consider the additional contribution to the gas
thermal energy given by the BH feedback, so that:
\begin{equation}
E_{th}(t)= E_{th,SN}(t)+E_{th,AGN}(t).
\end{equation}
The various simplifying assumptions adopted in the evaluation of the 
gas binding energy and of the BH feedback will be discussed in some detail 
in the following sections.

\subsection{The gas binding energy}

If we define
\begin{equation}
\Psi(r,t) = \pi R_e^2\psi(r,t)
\end{equation}
and
\begin{equation}
\dot{M}_{*} = \int_{0.1}^{80}\phi(M)\Psi(r,t-\tau_{M})R_{M}(t-\tau_{M})dM
\end{equation}
where $R_{M}$ is the return mass fraction (i.e. the fraction of mass
in a stellar generation that is ejected into the interstellar medium
by stars of mass $M$; see Tinsley, 1980)
and $\tau_{M}$ is the lifetime of a star of mass $M$, then, before the
galactic wind, the mass of gas in the bulge evolves at a rate
\begin{equation}
\dot{M}_{g}(t<t_{GW}) = \dot{M}_{inf} + \dot{M}_{*} -\Psi(r,t) - \dot{M}_{BH},
\end{equation}
i.e. gas is accreted from the halo, is restituted by stellar mass loss
and is subtracted by star formation and black hole accretion.  
After the wind, star formation vanishes and we suppose that, due to the
development of a global outflow, the infall is arrested. 
Therefore, this equation reduces to:
\begin{equation}
\dot{M}_{g}(t>t_{GW}) = \dot{M}_{*} - \dot{M}_{BH} - \dot{M}_{W}
\end{equation}
where $\dot{M}_{W}$ is the rate of mass loss due to the galactic wind.
We assume that all the gas present at a given time is lost, so that
$\dot{M}_{W} = \dot{M}_{*} - \dot{M}_{BH}$.
In the present treatment the galaxy mass model is required in
order to estimate the binding energy of the gas in the bulge, a key
ingredient in the establishment of the wind phase. In particular, the
current galaxy model is made by three different components, namely:

\begin{itemize}

\item[-]A spherical Hernquist (1990) distribution representing the stellar
component of the bulge:

\begin{equation}
\left\{
\begin{array}{lcl}
\rho_b(r) & = & {\displaystyle \frac{M_b}{2\pi}\frac{\rb}{r(r+\rb)^3}},\cr
\Phi_b(r) & = & -{\displaystyle \frac{GM_b}{r+\rb}},
\end{array}
\right.
\end{equation}
where $M_b$ is the bulge mass, $\Phi_b$ is the bulge potential and
$\rb$ is the scale radius of the bulge, related to its effective
radius by $R_e\simeq 1.8\rb$.

\noindent\item[-]A spherical 
isothermal dark matter halo with circular velocity $v_c$:
\begin{equation}
\left\{
\begin{array}{lcl}
\rho_{DM}(r) & = & {\displaystyle\frac{v_c^2}{4\pi Gr^2}}\cr
\Phi_{DM}(r) & = & {\displaystyle v_c^2\ln \frac{r}{r_0}}
\end{array}
\right.
\end{equation}
where $r_0$ is an arbitrary scale-length.

\noindent\item[-] A razor-thin exponential disk with surface density

\begin{equation}
\Sigma_d(R)=\frac{M_d}{2\pi\Rd^2}e^{-R/\Rd},
\end{equation}
where $M_d$ is the total disk mass, $R$ is the cylindrical radius, and
$\Rd$ is the disk scale radius.  As well known the gravitational
potential of a disk in general can be expressed by using the
Hankel-Fourier transforms (e.g., Binney \& Tremaine 1987): however, as
shown in Appendix~\ref{sec:A1}, under the assumption of a spherical
gas distribution, the contribution to the gas binding energy can be
easily calculated without using the explicit disk potential.

\end{itemize}

In fact, we assume that the gas distribution before the establishment
of the galactic wind is spherically symmetric, and parallel to the
stellar one, i.e.
\begin{equation}
\rho_g(r)=\frac{M_g}{2\pi}\frac{\rb}{r(r+\rb)^3}.
\end{equation}

In order to estimate the energy required to induce a bulge wind, we
define a {\it displacement radius} $\rt$, and we calculate the energy
required to displace at $\rt$ the gas contained at $r<\rt$ (while
maintaining spherical symmetry).  We adopt a fiducial value of $\rt
=3R_{e}$; the calculated values of the binding energy do not change
significantly for $\rt$ ranging from $2R_e$ to $10R_e$.  A spherically
symmetric displacement (while not fully justified theoretically)
allows a simple evaluation of the gas binding energy, and it is
acceptable in the present approach.  Consistently with the assumption
above,
\begin{equation}
\Delta\Eg = \Delta\Egb + \Delta\EgDM + \Delta\Egd ,
\end{equation}
where the various terms at the r.h.s. describe the gas-to-bulge,
gas-to-DM and gas-to-disk contributions.
Elementary integrations show that
\begin{eqnarray}
\Delta\Egb & = &
4\pi\int_0^{\rt}\rho_g(r)[\Phi_b(\rt)-\Phi_b(r)]r^2dr\cr
            &=& {GM_gM_b\over\rb} {\delta^3\over 3(1+\delta)^3}, 
\end{eqnarray}
and
\begin{eqnarray}
\Delta\EgDM&=& 
4\pi\int_0^{\rt}\rho_g(r)[\Phi_{DM}(\rt)-\Phi_{DM}(r)]r^2dr\cr
             &=& M_gv_c^2\left[\ln(1+\delta)-\frac{\delta}{1+\delta}\right],
\end{eqnarray}
where $\delta \equiv\rt/\rb$, while 
\begin{eqnarray}
\Delta\Egd = \frac{GM_gM_d}{\Rd} \times \Delta\widetilde{\Egd},
\end{eqnarray}
and the function $\Delta\widetilde{\Egd}$ is given in Appendix.
Of course, $\Delta\Eg$ is linearly proportional to the gas mass in the
bulge.

\begin{figure}
\includegraphics[width=0.5\textwidth]{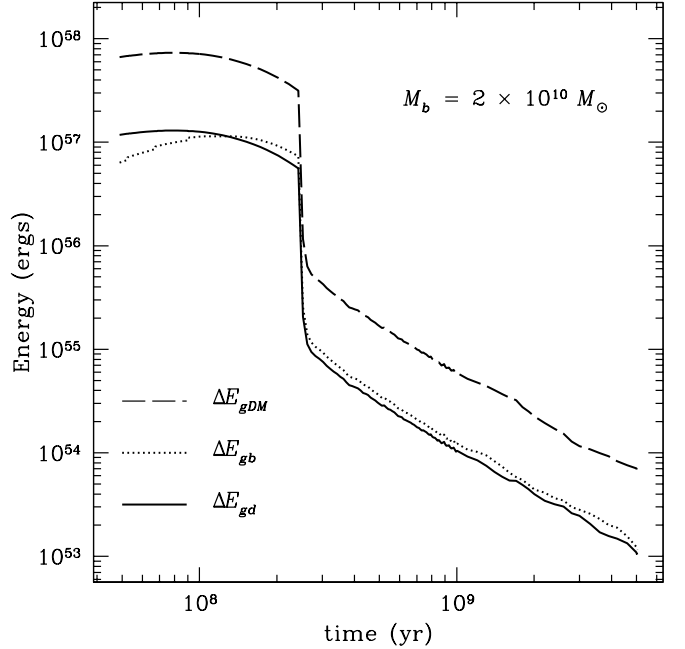}

\caption{Time evolution of the different contributions to the gas
binding energy in the bulge of a Milky Way like galaxy: dark matter
halo (dashed line), bulge (dotted line), and exponential disk (solid
line). In particular, $M_b=2\times10^{10}M_{\odot}$, 
$R_e=2$ kpc, $v_c=200$ km/s, $M_d=10^{11}M_{\odot}$ and $\Rd =4.3$
kpc.}
\label{fig:fig01} 
\end{figure}

In Fig. \ref{fig:fig01} we see that, for a $10^{10}M_{\odot}$ bulge
like ours, the dominant contribution arises from the dark matter halo,
whereas the bulge and disk contributions are comparable, both about
one order of magnitude smaller than the dark matter halo one.  This
differs from previous calculations (Ballero et al. 2007a) in the way
that the bulge contribution is reduced by almost one order of
magnitude.  The same is true for bulges of other masses.  We must
consider moreover that in the inside-out scenario for the Galaxy
formation (Chiappini et al. 1997) the disk will probably form much
later than the bulge, so its contribution to the potential well during
the bulge formation could be negligible.
Thus, what we explore is the extreme hypothesis that the disk has
  been in place since the beginning.

\subsection{Feedback from supernovae}

The cumulative thermal energy injected by SNe is calculated
as in Pipino et al. (2002).  Namely, if we call $R_{SN Ia/II}(t)$ the
rate of TypeIa/II SN explosions:
\begin{equation}
E_{th,SN}(t) = E_{th,SN Ia}(t) + E_{th,SN II}(t),
\end{equation}
where
\begin{equation}
E_{th,SN Ia/II}(t) = \int_0^t \epsilon(t-t')R_{SN Ia/II}(t')dt' \mbox{erg}.
\end{equation}

The evolution with time of the thermal content $\epsilon$ of a SN
remnant, needed in equation above, is given by (Cox, 1972):
\begin{equation}
\epsilon(t_{SN}) = \left\{
\begin{array}{llll}
7.2 \times 10^{50}\epsilon_0 &\mbox{erg} & \mbox{for} & 
0 \leq t_{SN} \leq t_c, \\
2.2 \times 10^{50}\epsilon_0 (t_{SN}/t_c)^{-0.62} &\mbox{erg} &
\mbox{for} & t_{SN} \geq t_c, 
\end{array}
\right.
\end{equation}
where $\epsilon_0$ is the initial blast wave energy of a SN in units
of $10^{51}$ erg, assumed equal for all SN types, $t_{SN}$ is the
time elapsed since explosion and $t_c$ is the metallicity-dependent
cooling time of a SN remnant (Cioffi et al., 1988):
\begin{equation}
t_c=1.49\times 10^4\epsilon_0^{3/14}n_0^{-4/7}\zeta^{-5/14} \mbox{ yr}.
\end{equation}
In this expression $\zeta=Z/Z_{\odot}$, and $n_0$ is the hydrogen number density.

\subsection{Accretion onto and feedback from the central black hole}

In our phenomenological treatment of BH feedback, we only considered
radiative feedback, thus neglecting other feedback mechanisms such as
radiation pressure and relativistic particles, as well as mechanical
phenomena associated with jets. From this point of view we are
following the approach described in Sazonov et al. (2005), even though
several aspects of the physics considered there (in the context of
elliptical galaxy formation) are not taken into account. In fact, we
note that these phenomena can be treated in the proper way only by
using hydrodynamical simulations.

We suppose that the bulge gas is fed into the
spherically accreting BH at the Bondi rate $\dot M_B$.  
However, the amount of accreting material cannot exceed the Eddington
limit, i.e.:
\begin{equation}
\dot{M}_{BH}=\min(\dot{M}_{Edd},\dot{M}_B)
\end{equation}


The Eddington accretion rate, i.e. the accretion rate beyond which
radiation pressure overwhelms gravity, is given by:
\begin{equation}
\dot{M}_{Edd}=\frac{L_{Edd}}{\eta c^2}
\label{eq:eq01}
\end{equation}
where $\eta$ is the efficiency mass-to-energy conversion. In
general, $0.001 \leq \eta \leq 0.1$; we adopt the maximum value
$\eta = 0.1$ (Yu \& Tremaine, 2002).
The Eddington luminosity is given by
\begin{equation}
L_{Edd}=1.3\times 10^{46}\frac{M_{BH}}{10^8M_{\odot}} \mbox{ erg s}^{-1}.
\end{equation}

The Bondi accretion rate describes the stationary flow of gas from
large distances onto the black hole, for a given gas temperature and
density (see Bondi, 1952), and is given by
\begin{equation}
\dot{M}_{B}=4\pi R_B^2\rho_Bc_S,
\label{eq:eq02}
\end{equation}
where
\begin{equation}
R_B=\frac{GM_{BH}\mu m_p}{2\gamma kT} = 16\mbox{ pc
}\frac{1}{\gamma}\frac{M_{BH}}{10^8M_{\odot}}
\left(\frac{T}{10^6\mbox{K}}\right)^{-1}
\end{equation}
with
\begin{equation}
c_s^2=\left(\frac{\partial p}{\partial\rho}\right)_{isot}=\frac{kT}{\mu m_p},
\end{equation}
and $\rho_B$ (the gas density at $R_B$) can be estimated as
\begin{equation}
\rho_B=\frac{\bar{\rho}_e}{3}\left(\frac{R_e}{R_B}\right)^2.
\end{equation}
In the code we adopt $\gamma = 1$ (isothermal flow).
If we assume that all the gas mass is contained within $2R_e$, the
mean gas density within $R_e$ is given by
\begin{equation}
\bar{\rho}_e=\frac{3M_g}{8\pi R_e^3}.
\end{equation}

The equilibrium gas temperature can be estimated as the bulge virial
temperature.
\begin{equation}
T_{vir}\simeq \frac{\mu m_p \sigma^2}{k} = 
3.0 \times 10^6 \mbox{ K } 
\left(\frac{\sigma}{200\mbox{ km s}^{-1}}\right)^2
\end{equation}
where $\sigma$ is the one-dimensional stellar velocity dispersion 
in the bulge, which is given by: 
\begin{equation}
\sigma^2\equiv -{\Wb\over 3M_b}.
\end{equation}
The virial potential trace $\Wb$ for the bulge is obtained by summing
the contribution of the three galaxy components, i.e.
\begin{eqnarray}
\Wb & = &  \Wbb +\WbDM + \Wbd\cr
    & = & - {GM_b^2\over 6\rb} -M_b v_c^2 - 
                 {GM_d M_b\widetilde{\Wbd}\over\Rd}
\end{eqnarray}
where the term $\widetilde{\Wbd}$ is given in Appendix.

The bolometric luminosity emitted by the accreting BH is then
calculated as: 
\begin{equation}
L_{bol}=\eta c^2\dot{M}_{BH}
\end{equation}

Finally, we assume that the energy released by the black hole is the
integral of a fraction $f$ of this luminosity over the time-step:

\begin{equation}
E_{th,BH}= f\int_{t}^{t+\Delta t} L_{bol}dt \simeq f L_{bol}\Delta t
\end{equation}
The value of $f$ can vary between 0 and 1; we assume $f=0.05$
(Di Matteo et al. 2005).

\begin{figure}
\centering
\includegraphics[width=0.5\textwidth]{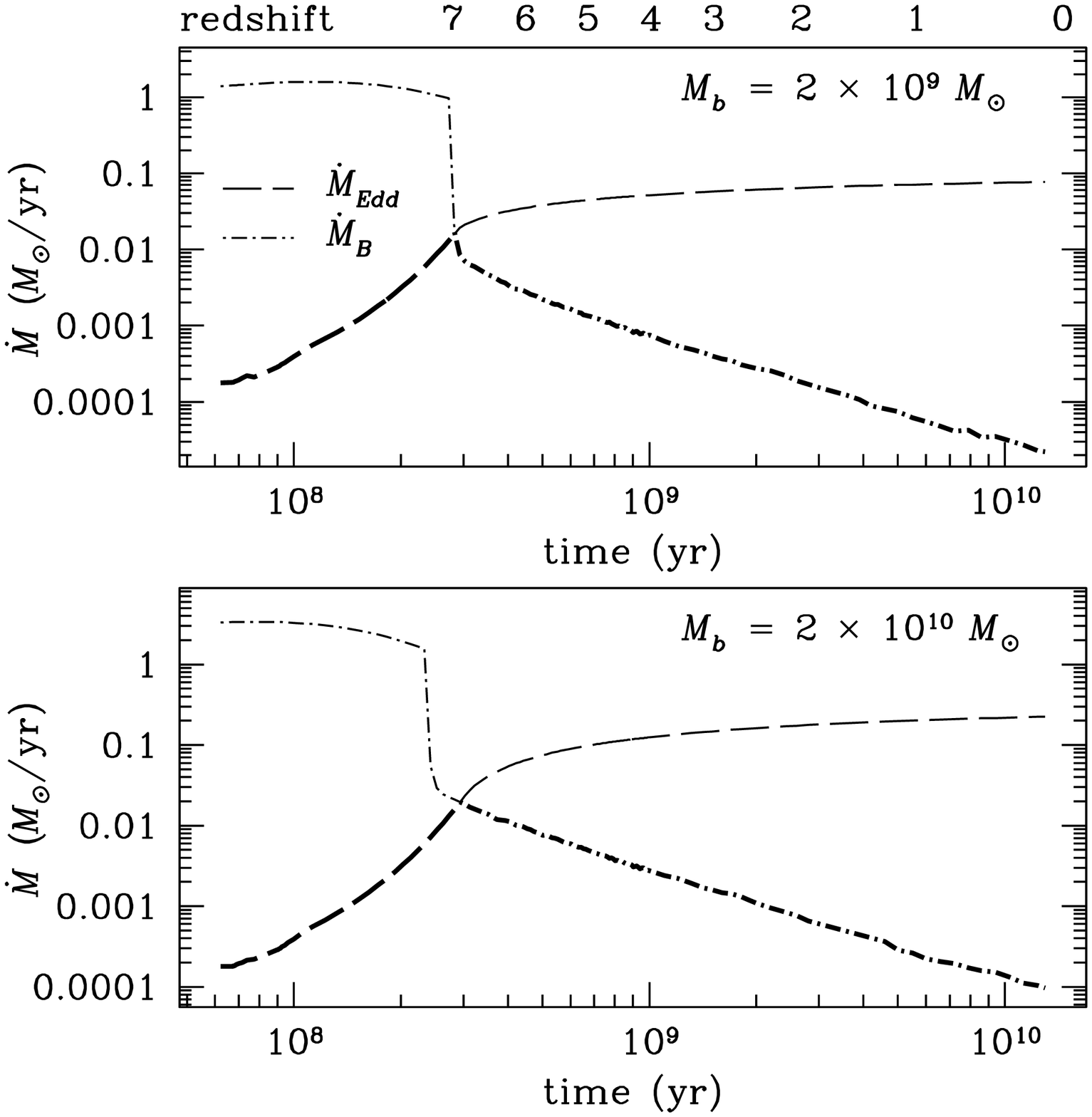}
\includegraphics[width=0.5\textwidth, clip, trim=0 275 0 30]{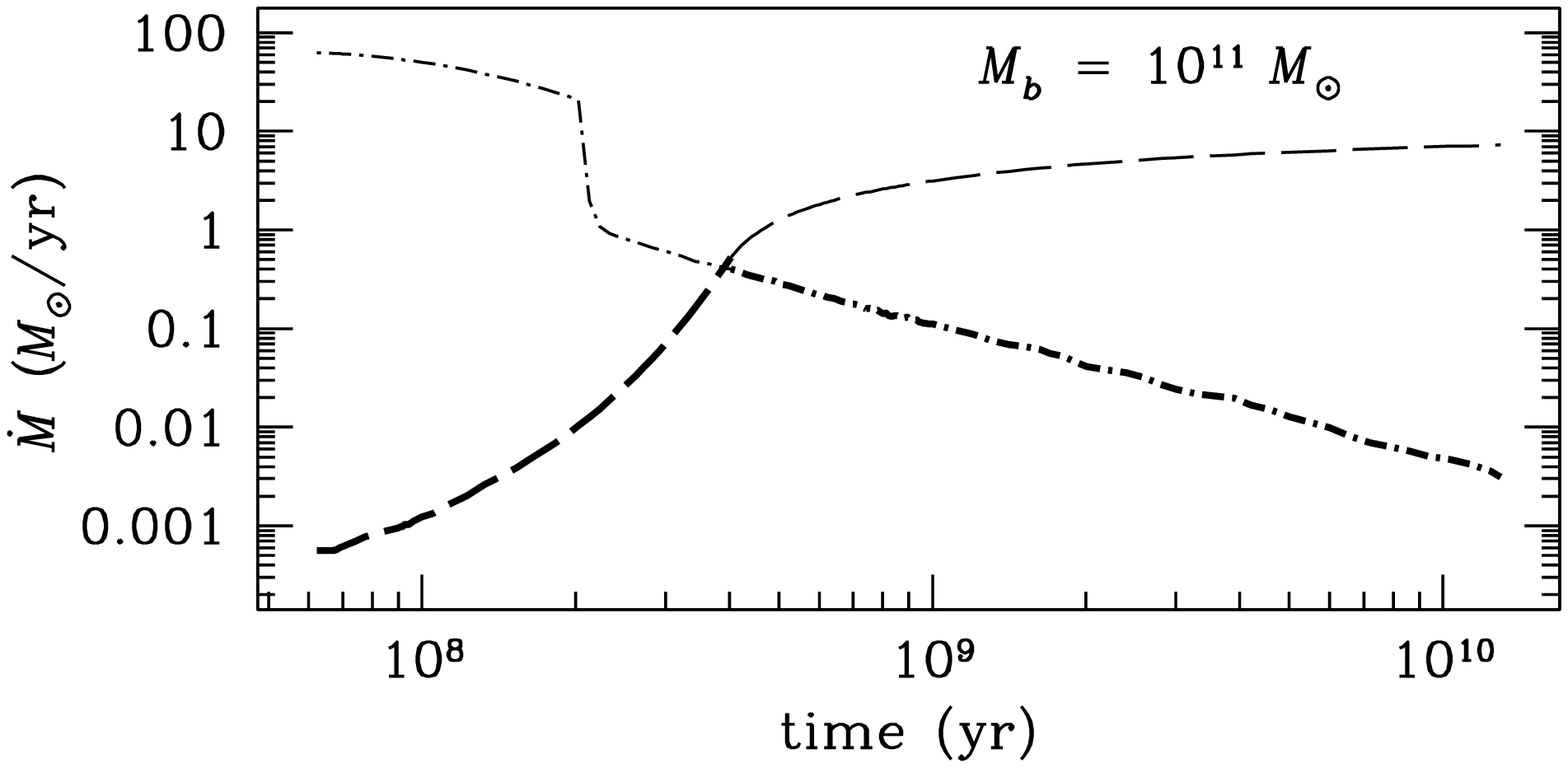}
\caption{Time evolution of the Eddington (dashed line) 
and Bondi (dot-dashed line) accretion rates for bulges of various masses. 
The thicker lines indicate the resulting accretion rate, which is
assumed to be the minimum between the two.}
\label{fig:fig02} 
\end{figure}
\begin{figure}
\centering
\includegraphics[width=0.5\textwidth]{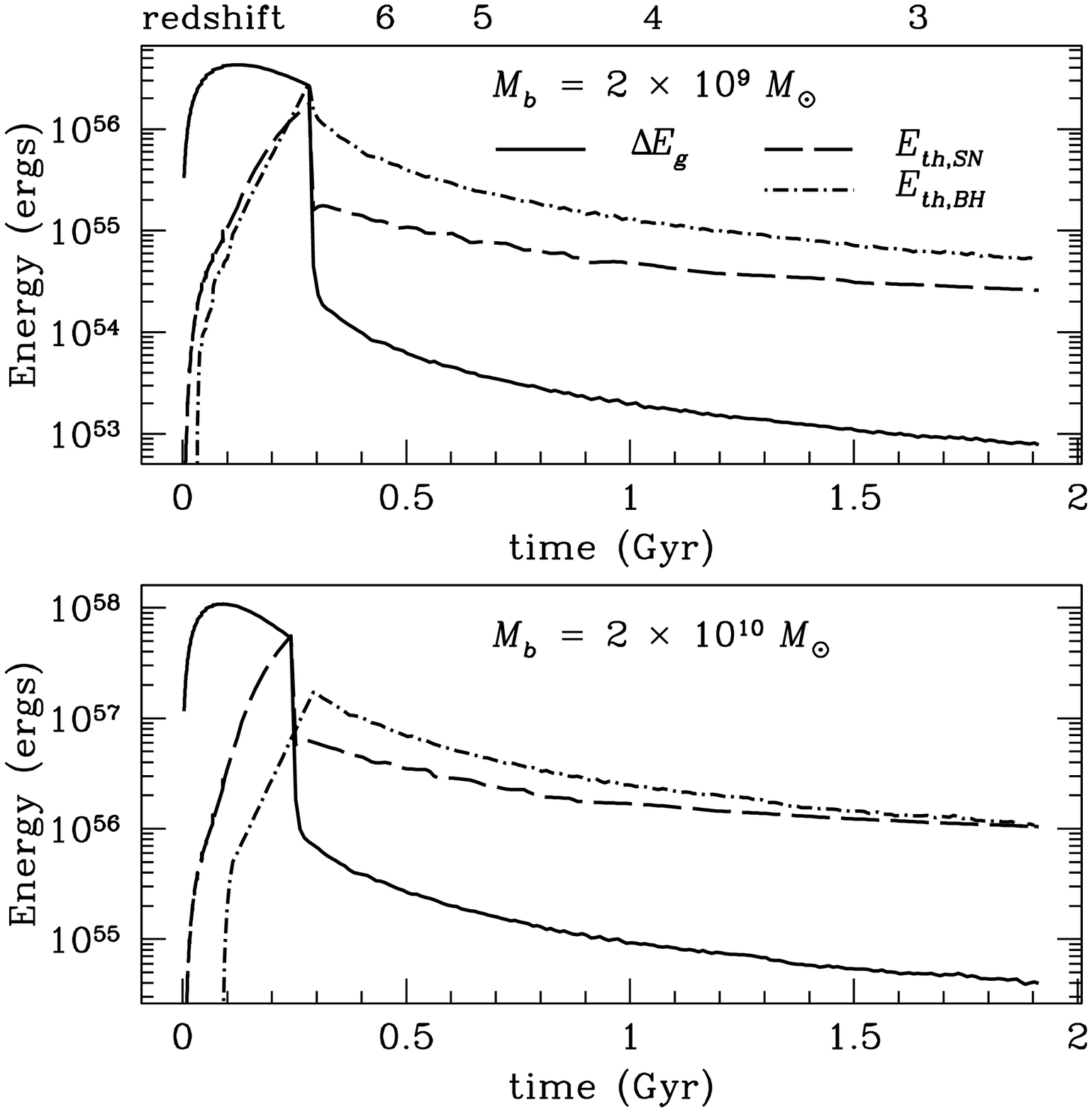}
\includegraphics[width=0.5\textwidth, clip, trim=0 275 0 30]{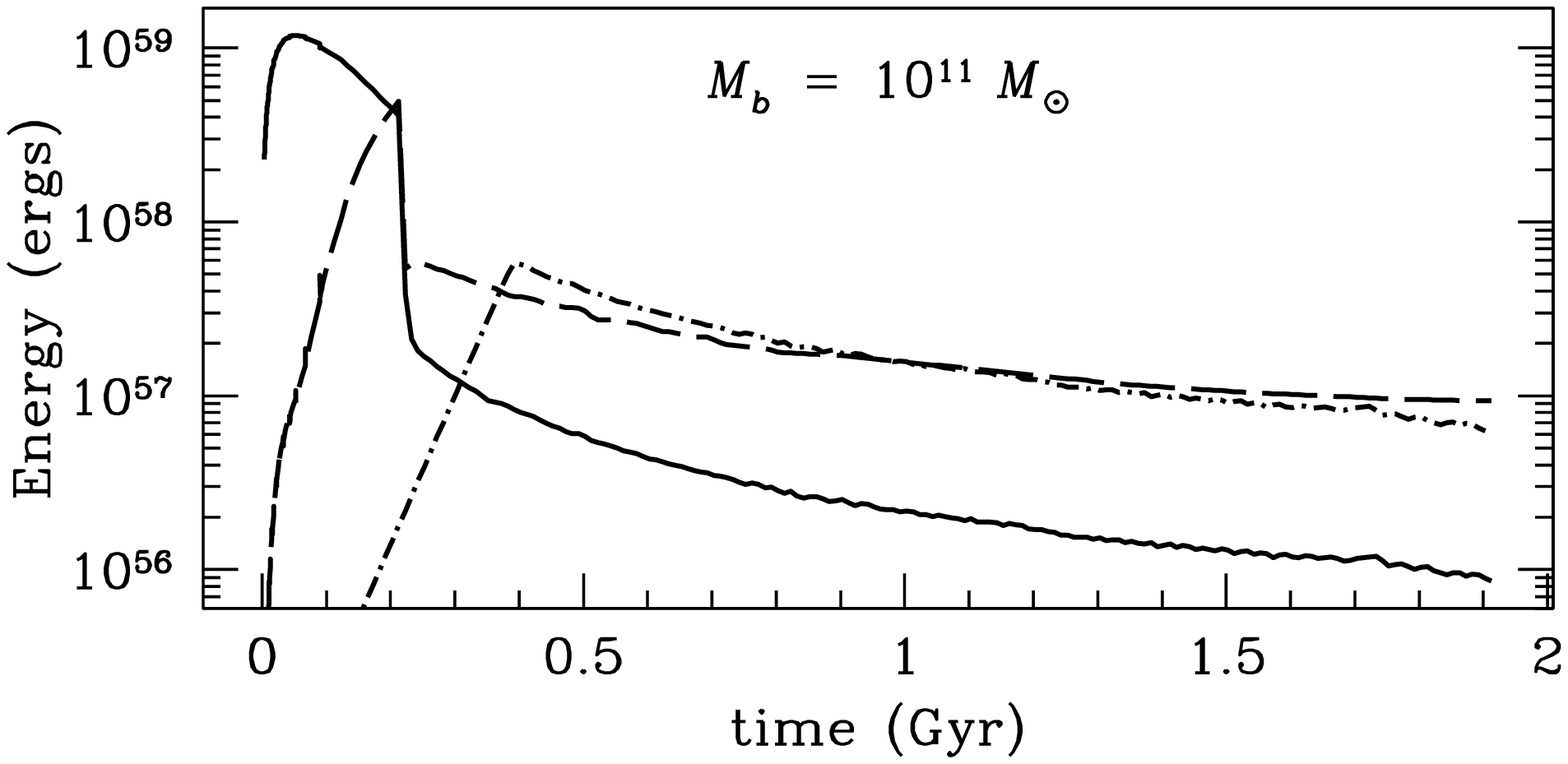}
\caption{Energy balance as a function of time for bulges of various
  masses. The figure shows the gas binding energy (solid line) compared
  to the thermal energy released by supernovae (dashed line) and the
  accreting black hole (dot-dashed line).}
\label{fig:fig03} 
\end{figure}

The seed black hole mass was modified in a range $5\times
10^{2}-5\times 10^{4}M_{\odot}$ without any appreciable change in the
results.  Therefore, we adopted a universal seed black hole mass of
$10^{3}M_{\odot}$.

\section{Results}

\subsection{Mass loss and energetics}

Fig. \ref{fig:fig02} shows the evolution of the
Eddington and Bondi accretion rates with time and redshift.
The redshift was calculated
assuming a $\Lambda$CDM cosmology with $H_{0} = 65$ km s$^{-1}$
Mpc$^{-1}$, $\Omega_{M} = 0.3$ and $\Omega_{\Lambda} = 0.7$, and a
redshift of formation of $z_{f}\simeq10$.   
We can see that the history of accretion onto the
central BH can be divided into two phases: the first,
Eddington-limited and the second, Bondi-limited.  
Most of the
accretion and fueling occurs around the period of transition between
the two phases, which approximately coincides with the occurrence of the
wind although it extends for some time further in the most massive
models.  
The details of the transition depend on the numerical treatment of the
wind, however since the gas consumption in the bulge is very fast, and
the Bondi rate depends on the gas density, we can expect that the
results would not change significanlty even if we 
modeled the galactic wind as a continuous wind.
The BH mass is essentially accreted, within a factor of two,
in a period ranging from 0.3 to 0.8 Gyrs, i.e. 2 to 6\% of the
bulge lifetime, which we assume to be 13.7 Gyr.

Fig. \ref{fig:fig03} compares the different contributions to the
thermal energy, i.e. the feedback from SNe and from the AGN, with the
potential energy.  We see that only in the case of $M_b =2 \times 10^9
M_{\odot}$ does the BH feedback provide a thermal energy comparable to
that produced by the SN explosions before the onset of the
wind\footnote{The Bondi accretion rate does not vanish even if the
thermal energy of the ISM is larger than the potential energy, due to
the fresh gas provided by the stellar mass losses of the aging stars
in the bulge.}. 
Therefore, in the context of chemical and photometrical evolution, the
contribution of the BH feedback is negligible in most cases, unless we
assume an unrealistically large fraction of the BH luminosity to be
transferred to the ISM. 
Note that this conclusion is also
supported by hydrodynamical simulations specifically designed to
study the effects of radiative BH feedback in elliptical galaxies
(Ciotti \& Ostriker 1997, 2001, 2007; Ostriker \& Ciotti 2005).
Also Di Matteo et al. (2003) showed that it is unlikely that 
black hole accretion plays a crucial role in the general process of
galaxy formation, unless there is strong energetic feedback by active
QSOs (e.g. in the form of radio~jets).

\subsection{Star formation rate}

\begin{figure}
\includegraphics[width=0.5\textwidth]{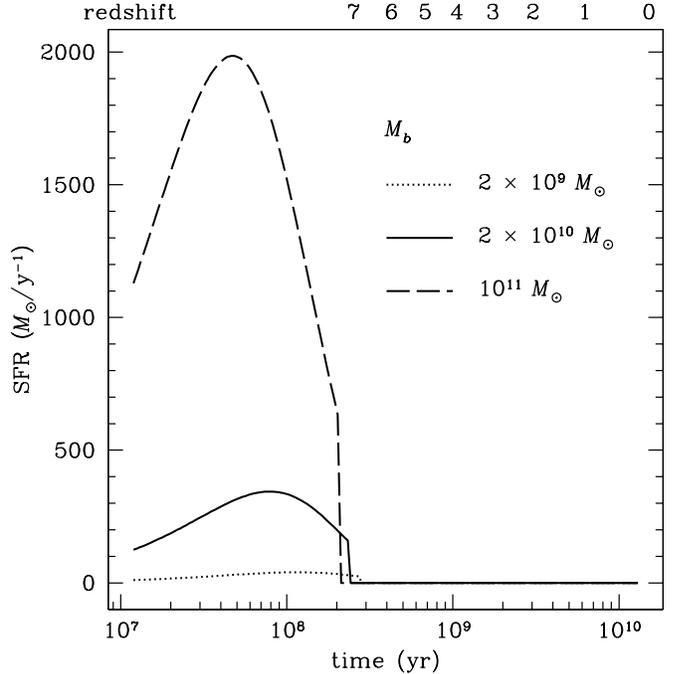}
\caption{Star formation rate as a function of time and 
redshift for bulges of various masses. The peak value of the 
$M_b=2\times 10^9 M_{\odot}$ case is $\sim 40 M_{\odot}$/yr.}
\label{fig:fig04} 
\end{figure}

Fig. \ref{fig:fig04} shows the evolution of the global SFR in the
bulge as a function of time and redshift.  The break corresponds to
the occurrence of the galactic wind.  
It is evident that in the
case of the most massive bulges it is easy to reach the very high SFRs
of a few times $1000 M_{\odot}$/yr inferred from observations at high
redshifts (e.g. Maiolino et al., 2005).  
It is worth noting that this results matches very well the statement
of Nagao et al. (2006) that the absence of a significant metallicity
variation up to $z \simeq 4.5$ implies that the active star-formation
epoch of QSO host galaxies occurred at $z \gtrsim 7$.
There is evidence for some tiny
downsizing (star formation ceases at slightly earlier times for larger
galaxies; see Table \ref{tab:tab1}).  
We stress that we are making predictions about single galaxies and not
about the AGN population, and we only want to show that it is possible
to achieve such high rates of star formation in a few Myrs.
 
\subsection{Black hole masses and luminosities}

\begin{figure}
\includegraphics[width=0.5\textwidth]{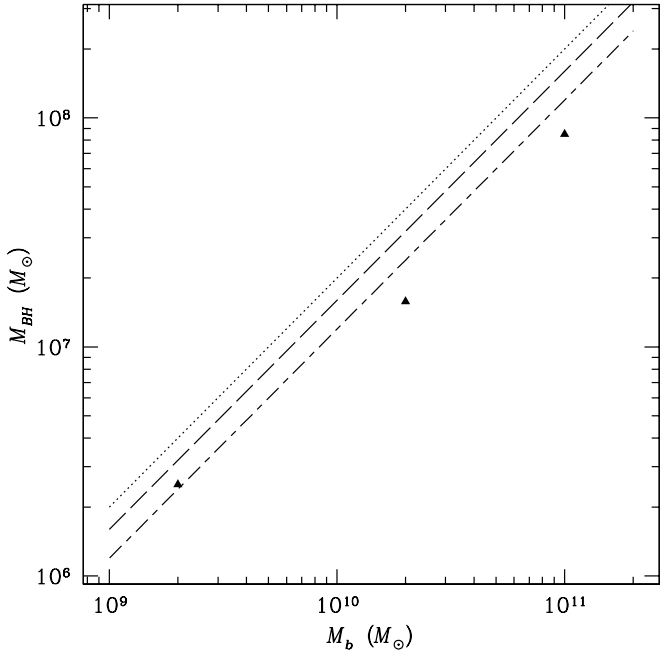}
\caption{Final black hole masses as a function of the bulge mass
  predicted by our models (triangles). The short dashed-long dashed
  line represents the measured relation of McLure \& Dunlop (2002),
  the dotted line the one of Marconi \& Hunt (2003) and the dashed
  line the one of H\"aring \& Rix (2004).}
\label{fig:fig05} 
\end{figure}

In Fig. \ref{fig:fig05} we show the final BH masses resulting from the
accretion as a function of the bulge mass.  The predicted BH masses,
which are reached in a few hundred Myrs, are in good agreement with
measurements of BH masses inside Seyfert galaxies (Wandel et al.,
1999; Kaspi et al., 2000; Peterson, 2003).  This is a valuable result,
given the simplistic assumptions of our model, since in this case we
did not have to stop accretion in an artificious way as in Padovani \&
Matteucci (1993); in fact, the BH growth at late times is limited
by the available amount of gas, as described by Bondi accretion.  
The figure suggests an approximately linear relation between the bulge and
BH mass, as first measured by Kormendy \& Richstone (1995) and
Magorrian et al. (1998).  Therefore, Seyfert galaxies appear to obey
the same relationship of quiescent galaxies and QSOs, as already
stated observationally by Nelson et al. (2004).

There were claims for a non-linear relation between spheroid and
black hole mass (Laor, 2001; Wu \& Han, 2001); however, measurements
of Marconi \& Hunt (2003) re-established the direct proportionality
between the spheroid mass $M_{sph}$ and $M_{BH}$: 
\begin{equation}
M_{BH} \simeq 0.0022 M_{sph}
\end{equation}
also in good agreement with recent
estimates from McLure \& Dunlop (2002) and Dunlop et al. (2003):
\begin{equation}
M_{BH} \simeq 0.0012 M_{sph}
\end{equation}
and from H\"aring \& Rix (2004):
\begin{equation}
M_{BH} \simeq 0.0016 M_{sph}
\end{equation}
The three relations of McLure \& Dunlop (2002), Marconi \& Hunt (2003)
and H\"aring \& Rix (2004)
are thus reported in the figure for comparison.
The agreement between measurements and predictions
is rather~good (within a factor of two).

\begin{figure}
\includegraphics[width=0.5\textwidth]{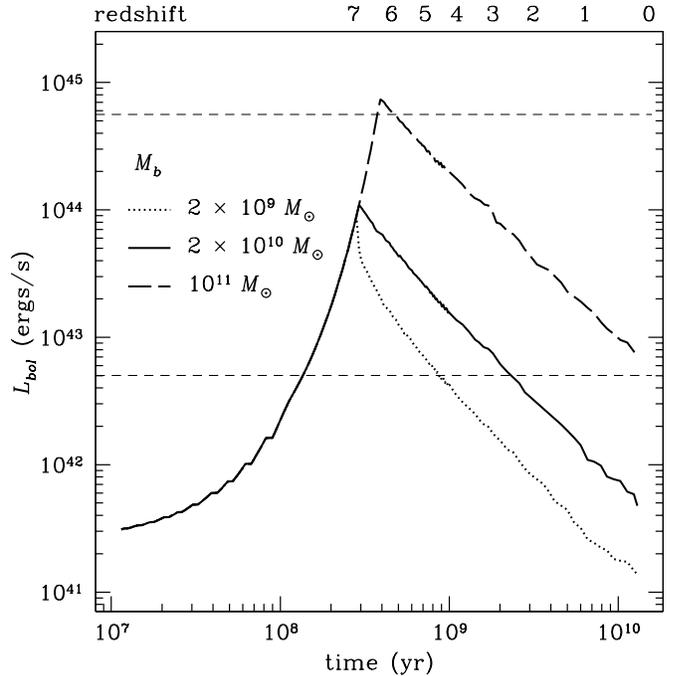}
\caption{Evolution with time and redshift of the bolometric 
luminosity emitted by the accreting BH for bulges of various masses. 
The dotted lines represent the range spanned by observations 
(see text for details).}
\label{fig:fig06} 
\end{figure}

In Fig. \ref{fig:fig06} we plot the predicted bolometric nuclear
luminosities $L_{bol}$ for the various masses, compared with
luminosity estimates for the Seyfert population (see e.g. Gonz\'alez
Delgado et al., 1998; Markowitz et al., 2003; Brandt \& Hasinger,
2005; Wang \& Wu, 2005; Gu et al., 2006).  Of course, $L_{bol}$ is
proportional to the calculated mass accretion rate
(Fig. \ref{fig:fig02}).  
In the first part, the plots overlap because
the Eddington luminosity depends only on the BH mass, and it is
independent on the galaxy mass; on the contrary, the Bondi accretion
rate is sensitive on the galaxy features.  
The break in the plot
corresponds to the time when the Bondi accretion rate becomes smaller
than the Eddington accretion rate (see Eqs. \ref{eq:eq01} and
\ref{eq:eq02}), and the break occurs later and later for more massive
galaxies, which therefore keep accreting at the Eddington rate for
longer timescales.  
This helps explaining why the outcoming final
black hole mass is proportional to the adopted bulge mass.

We see that the luminosities near the maximum are reproduced by models
of all masses. 
The same is not true for local ($z\simeq0$) Seyferts;
only the most massive model yields a bolometric luminosity which lies
in the observed range while for other masses the disagreement reaches
a factor of $\sim100$.
To simply shift the epoch of star formation of the less massive
Seyferts, would require unrealistically young bulges ($\lesssim 1$
Gyr), whereas one of our main assumptions is that spiral bulges are
old systems.
Padovani \& Matteucci (1993) assumed that the BH accreted the whole
mass lost from evolved stars.
If we calculate the fraction $\dot{M_B}/\dot{M_*}$ at any given time,
we see that its value is very close to $0.01$.
This explains why they obtained the correct nuclear bolometric luminosity
of local radio-loud QSOs, but severely overestimated the mass of the
resulting BH, which led them to suggest that the accretion phase
should last for a period not longer than a few $10^8$ years, at
variance with the present work. 
Moreover, it is not physically justified to assume that all the mass
expelled from dying stars falls onto the black hole.

One way to overcome this problem is to suppose that the smallest
bulges undergo a rejuvenation phase, which is possible because spiral
bulges are not isolated systems, but an interaction with their
surrounding disks can be triggered in several ways (bar
instabilities, minor mergers, fly-by's, and so on).
Thomas \& Davies (2006), using models combining recent star formation
with a base old population, found that the smallest bulges must have
experienced star formation events involving $10-30$\% of their total
mass in the past $1-2$ Gyr.
Such episodes may help explain as well the presence of the Galactic
bulge Mira population, whose calculated age is $1-3$ Gyr (van
Loon et al., 2003; Groenewegen \& Blommaert, 2005), and of young star
and star clusters in the very center of the Galaxy (Figer et al.,
1999; Genzel et al., 2003).
A small fraction of intermediate-age stellar population was also
detected in Seyfert 2 nuclei (Sarzi et al.,~2007).

Finally, we remark that by means of the present (non-hydrodynamical) modeling
of the BH accretion rate and of the gas budget, a truly realistic
accretion cannot be reproduced, whereas in a more realistic treatment
the AGN feedback would cause the luminosity to switch on and off
several times at the peaks of luminosity.  
We also note that at late times, when the accretion is
significantly sub-Eddington, a considerable reduction in the emitted
AGN luminosity might result as a consequence of a possible
radiatively inefficient accretion mode (e.g., Narayan \& Yi
1994). This could reproduce the quiescence of black holes at the
centre of present-day inactive spiral galaxies. 

\subsection{Chemical abundances and metallicities}

\subsubsection*{Observations for QSOs and Seyferts}

Estimates of chemical abundance ratios in AGNs are not an easy task.
The use of emission lines is subject to large uncertainties due to the
dependence of the lines on several parameters which are difficult to
quantify (e.g. column density, microtubulence, collisional excitation,
and so on).  This is a problem since most measurements of the [Fe/Mg]
abundance ratio, which provides a clock for constraining the ages of
QSOs and their timescales of enrichment, rely on the flux ratio of the
Fe {\scshape ii} (UV bump) and Mg {\scshape ii} ($\lambda2800$)
emission lines.  The true physical origin of the Fe {\scshape ii} UV
complex was put in question by Verner et al. (2003), Baldwin et
al. (2004) and Korista et al. (2004), who showed that the flux ratio
does not scale directly with the abundance ratio due to the
thermostatic effect of coolants.  Absorption lines could be a better
probe for QSO environments, especially narrow absorption lines which
avoid saturation and blending of important abundance diagnostics.
However, these data require large signal-to-noise ratios and are
therefore harder to obtain.  Moreover, they are not free from
uncertainties, concerning the shape of ionizing spectrum, the lack of
ionization constraints, the unknown coverage fraction and the exact
location of the absorbers (see Hamann \& Ferland, 1999 for an
extensive review about QSO abundance diagnostics).

Given these caveats, it is evident in any case that the emission
spectrum of AGNs is particularly alike for a very large range of
redshifts and luminosities (Osmer \& Shields, 1999).  The most
probable explanation is a similarity of chemical abundances.  In
particular, the analysis of the Fe {\scshape ii}(UV bump)/Mg {\scshape
ii}($\lambda2800$) flux ratio in various redshift ranges (Thompson et
al., 1999; Iwamuro et al., 2002; Freudling et al., 2003; Dietrich et
al., 2003a; Barth et al., 2003; Maiolino et al., 2003; Iwamoto et al.,
2004) was consistent with an [Fe/Mg] abundance ratio slightly
supersolar and almost constant for redshifts out to $z \sim 6.4$; 
this result was also confirmed for $z \sim 6$ QSOs by very recent 
works (Kurk et al., 2007; Jiang et al., 2007).
A weak trend with luminosity was detected (Dietrich et al., 2003a). 

Due to the time delay necessary for Fe enrichment from Type~II SNe in a
star formation history typical of elliptical galaxies (see Matteucci
\& Recchi, 2001), this means that the surrounding stellar population
must be already in place by the time the AGN shines and that, in a
well-mixed ISM, star formation must have begun $\gtrsim 10^{8}$ years
before the observed activity, i.e. at redshifts $z_{f} \gtrsim 10$
(Hamann et al., 2002; Hamann et al., 2004).

Many efforts were devoted to the measurement of nitrogen lines, since
due to its secondary nature\footnote{We recall that an element is
secondary when it is produced starting from a seed non-primordial
nucleus; therefore, its abundance is very sensitive on the metallicity
of the gas where it is formed.}, the N abundance relative to its seed
nuclei (e.g. oxygen) is a good proxy for metallicity.  As abundance
indicators, broad emission line ratios of N~{\scshape v}/C~{\scshape
iv} and N~{\scshape v}/He~{\scshape ii} were analysed by Hamann \&
Ferland (1993). Hamann \& Ferland (1999) also indicated N~{\scshape
v}/C~{\scshape iv} and N~{\scshape v}/O~{\scshape vi} in narrow
absorption line systems as possible abundance indicators.  
Hamann et al. (2002) instead favored N~{\scshape iii}]/O~{\scshape
iii}] and N~{\scshape v}/(C~{\scshape iv} + O~{\scshape vi}) as the
most robust abundance indicators.  The general conclusions drawn from
these diagnostics are that the AGNs appear to be metal rich at all
observed redshifts, with metallicities ranging from solar up to
$\sim10$ times solar. 
Dietrich et al. (2003b) studied a sample of 70 high-redshift QSOs
($3.5 \lesssim z \lesssim 5.0$) and based on emission-line flux
ratios involving C, N, O and He they estimated an average overall
metallicity of $\sim$ 4$-$5 $Z_{\odot}$ for the emitting gas.
A similar estimate ($\sim$5 $Z_{\odot}$) was drawn more recently by
Nagao et al. (2006) who examined 5344 spectra of high-redshift ($z
\geq 2$) QSOs taken from the SDSS DR2; they also confirmed the 
detected trend in the N/He and N/C emission line ratios which suggests
that there might be a correlation with luminosity, 
i.e. more luminous QSOs, residing in more massive galaxies, are more
metal rich.  Bentz et al. (2004) suggested that some very
nitrogen-enriched QSOs could be viewed at the peak of metal
enrichment, e.g. near the end of their accretion phases, although
their conclusions are not firm.  Work carried out on intrinsic narrow
absorption line systems (Petitjean \& Srianand, 1999; Hamann et al.,
2003; D'Odorico et al. 2004) confirmed that the observed N, C and Si
abundance ratios are consistent with at least solar metallicities.  In
particular, D'Odorico et al. (2004) interpreted their observations as
suggestive of a scenario of rapid enrichment due to a short ($\sim 1$
Gyr) star formation burst.  The amount of emission from dust and CO in
high-redshift QSOs (Cox et al., 2002) corroborate the idea of massive
amounts of star formation preceding the shining of the AGN.  However,
there is no need for exotic scenarios to explain the production of
heavy elements near QSOs (e.g. central star clusters, star formation
inside accretion disks), since normal chemical evolution of
ellipticals is sufficient to this purpose (Hamann \& Ferland, 1993;
Matteucci \& Padovani, 1993).

The majority of conclusions concerning QSOs has been found to hold
also for Seyfert galaxies, i.e. observations seem to confirm that most
Seyfert galaxies are metal rich.

An overabundance of nitrogen of a factor ranging from about 2 to 5 was
first detected in the narrow line region of Seyferts by
Storchi-Bergmann \& Pastoriza (1989, 1990), Storchi-Bergmann et
al. (1990) and Storchi-Bergmann (1991), and was later confirmed by
Schmitt et al. (1994).  Further work by Storchi-Bergmann et al. (1996)
allowed to derive the chemical composition of the circumnuclear gas in
11 AGNs, and high metallicities were found (O ranging from solar to
$2-3$ times solar and N up to $4-5$ times solar).  This trend was also
measured in more recent works (Wills et al., 2000; Mathur,
2000). Fraquelli \& Storchi-Bergmann (2003) examined the extended
emission line region of 18 Seyferts and claimed that the range in the
observed [N {\scshape ii}]/[O {\scshape ii}] line ratios can only be
reproduced by a range of oxygen abundances going from 0.5 to 3 times
solar.  Rodr{\'i}guez-Ardila et al. (2005), by a means of a
multi-cloud model, deduced that a nitrogen abundance higher than solar
by a factor of at least two would be in agreement with the [N
{\scshape ii}]+/[O {\scshape iii}]+ line ratio observed in the narrow
line Seyfert 1 galaxy Mrk 766.  Finally, Fields et al. (2005a), using
a simple photoionization model of the absorbing gas, found that the
strongest absorption system of the narrow line Seyfert 1 (NLS1) galaxy
Mrk 1044 has N/C $\gtrsim 4$(N/C)$_{\odot}$.

In the circumnuclear gas of the same galaxy, using column density
measurements of O {\scshape vi}, C {\scshape iv}, N {\scshape v} and H
{\scshape i}, Fields et al. (2005b) claimed that the metallicity is
about 5 times solar.  This is consistent with expectations from
previous studies: Komossa \& Mathur (2001), studying the influence of
metallicity on the multi-phase equilibrium in photoionized gas, stated
that in objects with steep X-ray spectra, such as NLS1s, such an
equilibrium is not possible if $Z$ is not supersolar.  Studying
forbidden emission lines, Nagao et al. (2002) derived $Z \gtrsim 2.5
Z_{\odot}$ in NLS1, whereas the gas of broad line Seyferts tends to be
slightly less metal rich.

An overabundance of iron was suggested to explain the strong optical
Fe {\scshape ii} emission in narrow line Seyferts (Collin \& Joly,
2000) and could provide an explanation for the absorption features
around $\sim1$ keV seen in some of these galaxies (Ulrich et al.,
1999; Turner et al., 1999) or for the strength of the FeK$\alpha$
lines (Fabian \& Iwasawa, 2000).  Lee et al. (1999), by constraining
the relationship between iron abundance and reflection fraction,
showed that the observed strong iron line intensity in the Seyfert
galaxy MCG$-$6-30-15 is explained by an iron overabundance by a factor
of $\sim2$ in the accretion disk.  Ivanov et al. (2003) found values
for [Fe/H] derived from the Mg {\scshape i} 1.50$\mu$m line ranging
from $-0.32$ to $+0.49$, but these values were not corrected for
dilution effects from the dusty torus continuum, and therefore are
probably underestimated.

\begin{table}
\centering
\footnotesize
\begin{tabular}{lrrr}
\hline
\hline
$M_{b}$ ($M_{\odot}$)& $2\times 10^{9}$ & $2\times 10^{10}$ & $10^{11}$ \\
\hline
\hline
$Z/Z_{\odot}$	& $ 4.23$	& $ 6.11$	& $ 7.22$	\\
$[$Fe/H$]$      	& $+0.83$	& $+0.96$	& $+1.07$	\\
\hline
$[$O/H$]$		& $+0.18$	& $+0.36$	& $+0.48$	\\
$[$Mg/H$]$		& $+0.38$	& $+0.55$	& $+0.69$	\\
$[$Si/H$]$		& $+0.79$	& $+0.98$	& $+1.13$	\\
$[$Ca/H$]$		& $+0.63$	& $+0.88$	& $+1.06$	\\
\hline
$[$O/Fe$]$		& $-0.68$	& $-0.60$	& $-0.59$	\\
$[$Mg/Fe$]$		& $-0.45	$	& $-0.41	$	& $-0.38	$	\\
$[$Si/Fe$]$		& $-0.04	$	& $+0.02	$	& $+0.06	$	\\
$[$Ca/Fe$]$		& $-0.19$	& $-0.07$	& $+0.00$	\\
\hline
$[$N/H$]$		& $+0.28$	& $+0.63$	& $+0.87$	\\
$[$C/H$]$		& $+0.15$	& $+0.24$	& $+0.30$	\\
\hline
$[$N/C$]$		& $+0.13$	& $+0.40$	& $+0.57$	\\
\hline
$t(Z_{\odot})$ (yr) & $10^{8}$ & $6\times10^{7}$ & $3\times10^{7}$\\
\hline
\end{tabular}
\caption{Values assumed by the chemical abundance ratios and by the metallicity in solar units after the wind by bulges of different masses. The table also reports the time by which $Z_{\odot}$ is reached in the different models (last line).}
\label{tab:tab2}
\end{table}

\subsubsection*{Metallicity and elemental abundances}

The mean values attained by the metallicity and the examined abundance
ratios for $z\lesssim6$ are resumed in Table
\ref{tab:tab2}\footnote{We recall that throughout this paper we use the
standard notation, i.e. for example [Fe/H] =
$\log($Fe/H$)-\log($Fe/H$)_{\odot}$.}.  Note that all the
[$\alpha$/Fe] ratios are undersolar or solar.

\begin{figure}
\centering
\includegraphics[width=0.5\textwidth]{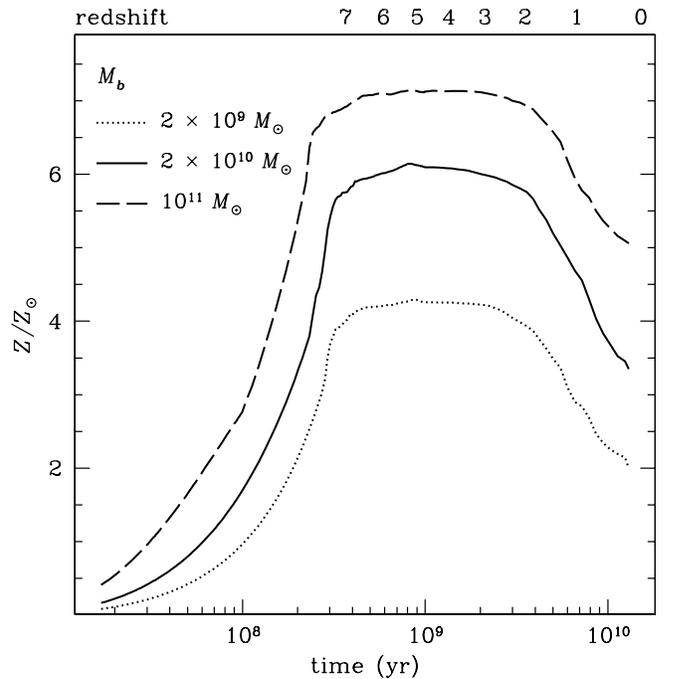}
\caption{Evolution with time (bottom axis) and redshift (top axis) of
  the metallicity in solar units $Z/Z_{\odot}$. The different curves
  denote different bulge masses as indicated in the upper left
  corner.} 
\label{fig:fig07} 
\end{figure}

Fig. \ref{fig:fig07} shows the evolution with time and redshift of the
metallicity $Z$ in solar units, for a standard $\Lambda$CDM scenario
with $H_{0}=65$ km s$^{-1}$ Mpc$^{-1}$, $\Omega_{M}=0.3$ and
$\Omega_{\Lambda} = 0.7$.

It can be seen that solar metallicities are reached in a very short
time, ranging from about $3\times 10^{7}$ to $10^{8}$ years with
decreasing bulge mass (see Table \ref{tab:tab2}).  
We notice that due to the $\alpha$-enhancement typical of spheroids,
solar $Z$ is always reached well before solar [Fe/H] is attained,
which occurs at times $\sim(1-3)\times10^{8}$ years.  
Then, $Z$ remains approximately
constant for the contribution of Type Ia SNe and low-and
intermediate-mass stars, before declining in the very late phases.
The very high metallicities inferred from observations
(e.g. Hamann et al, 2002; Dietrich et al., 2003b, their Figures 5 and
6) are thus very easily achieved.
More massive bulges give rise to higher metallicities, which is in
agreement with the statement that more luminous AGNs are more metal
rich, if we assume that more massive galaxies are also more luminous.
This assumption is supported observationally by e.g. Warner et
al. (2003) who found a positive correlation between the mass of the
supermassive BH and the metallicity derived from emission lines
involving N {\scshape v} in 578 AGNs spanning a large range in
redshifts.

\begin{figure}
\centering
\includegraphics[width=0.5\textwidth]{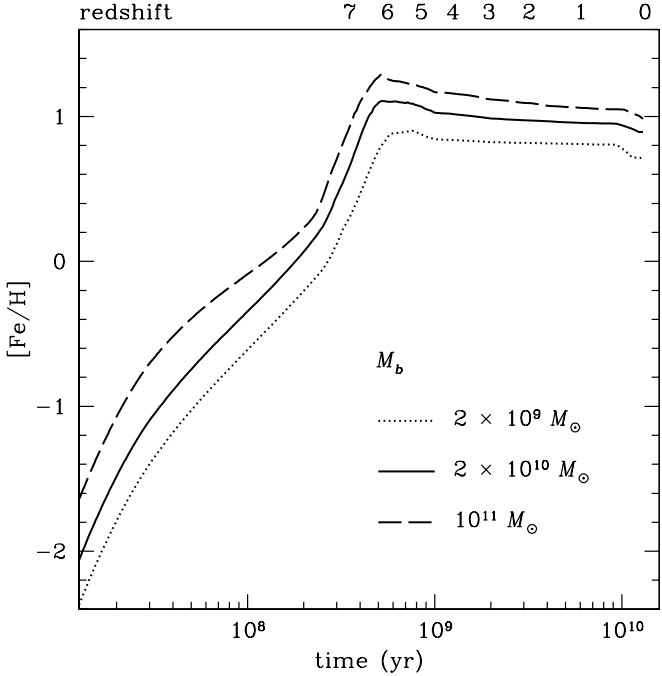}
\caption{Evolution with time (bottom axis) and redshift (top axis) of
  the [Fe/H] abundance ratio in bulges of various masses, as indicated
  in the lower right corner.} 
\label{fig:fig08} 
\end{figure}

\begin{figure*}
\centering
\includegraphics[width=0.31\textheight]{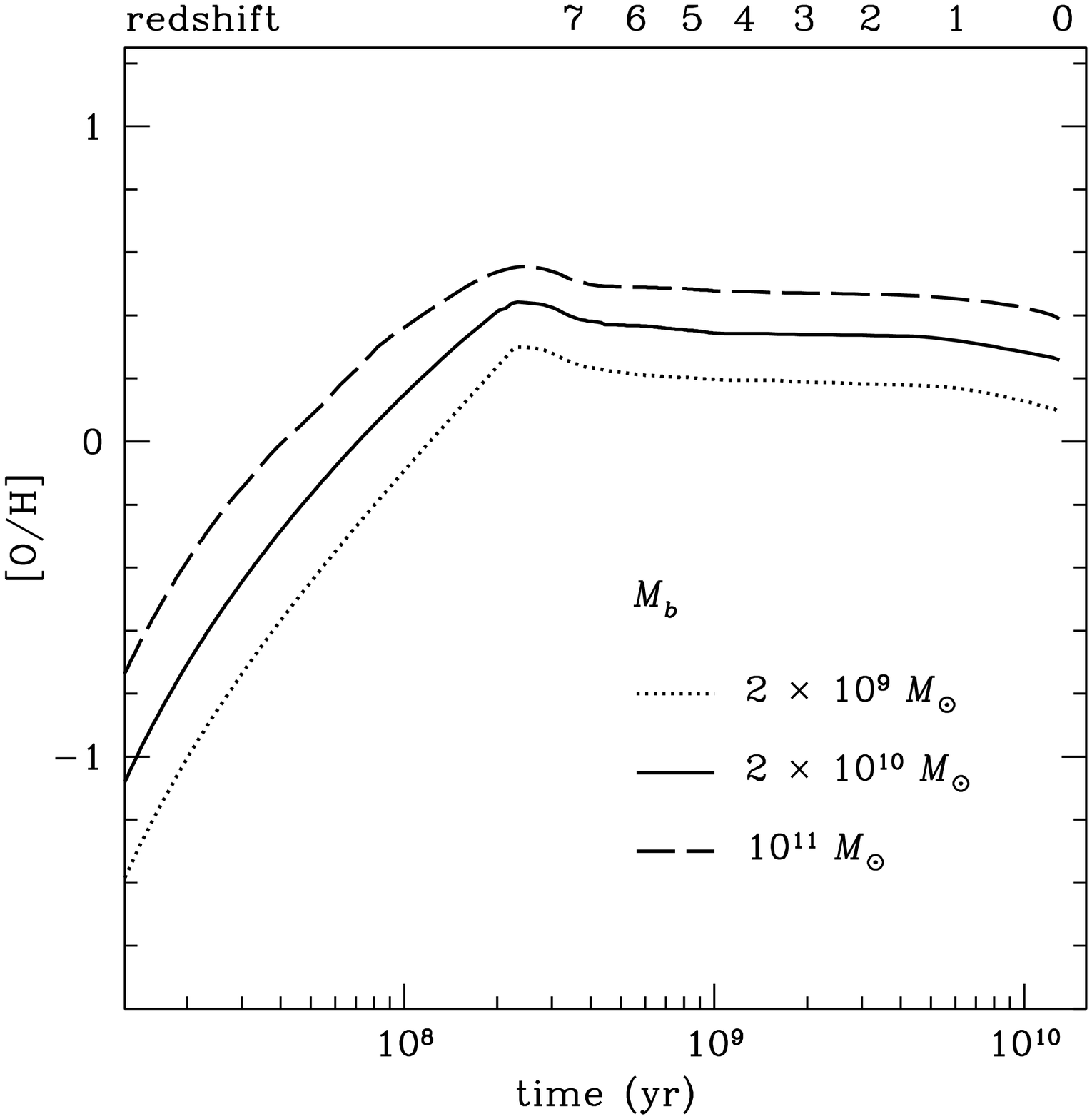}%
\includegraphics[width=0.31\textheight]{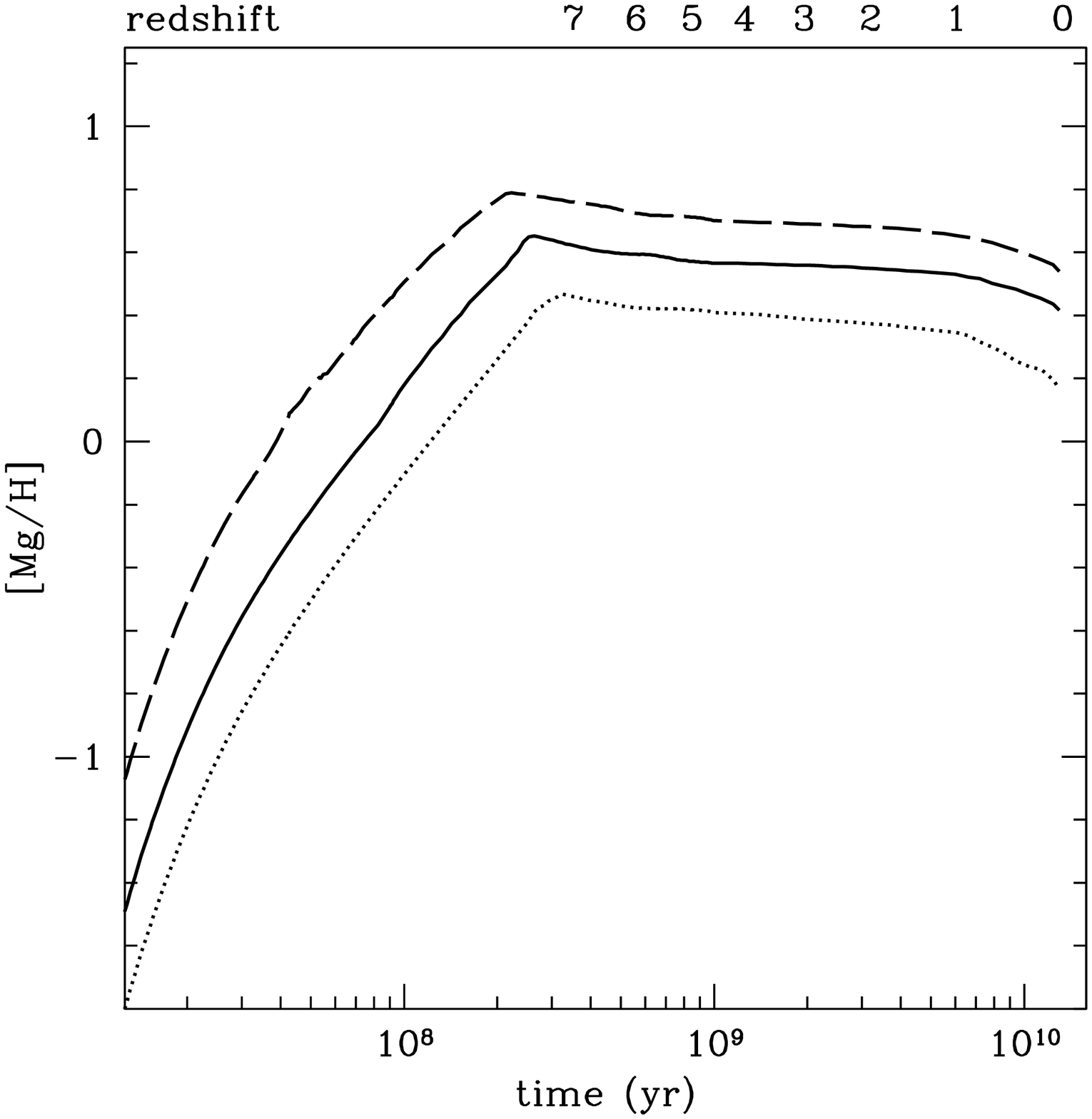}
\includegraphics[width=0.31\textheight]{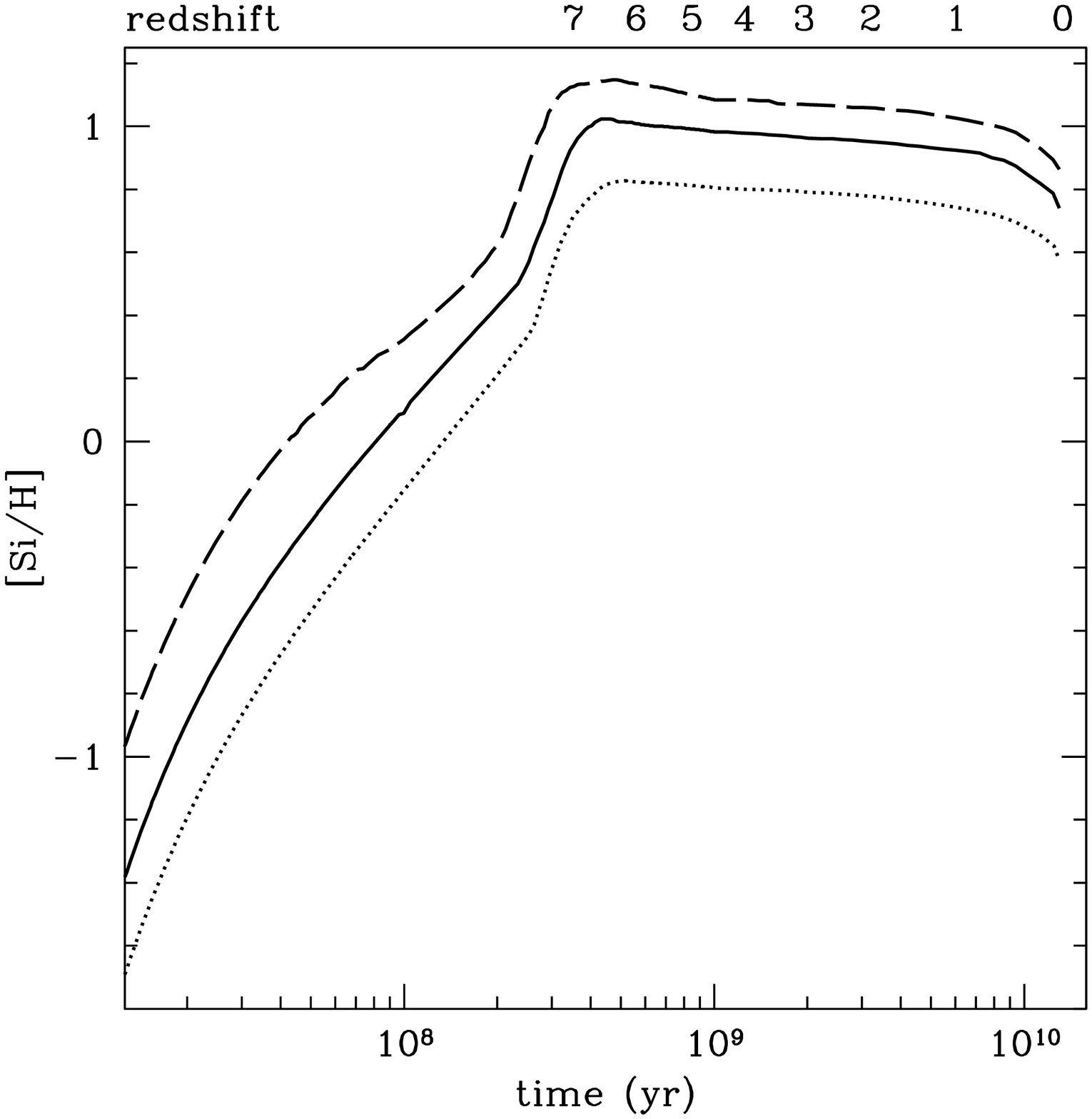}%
\includegraphics[width=0.31\textheight]{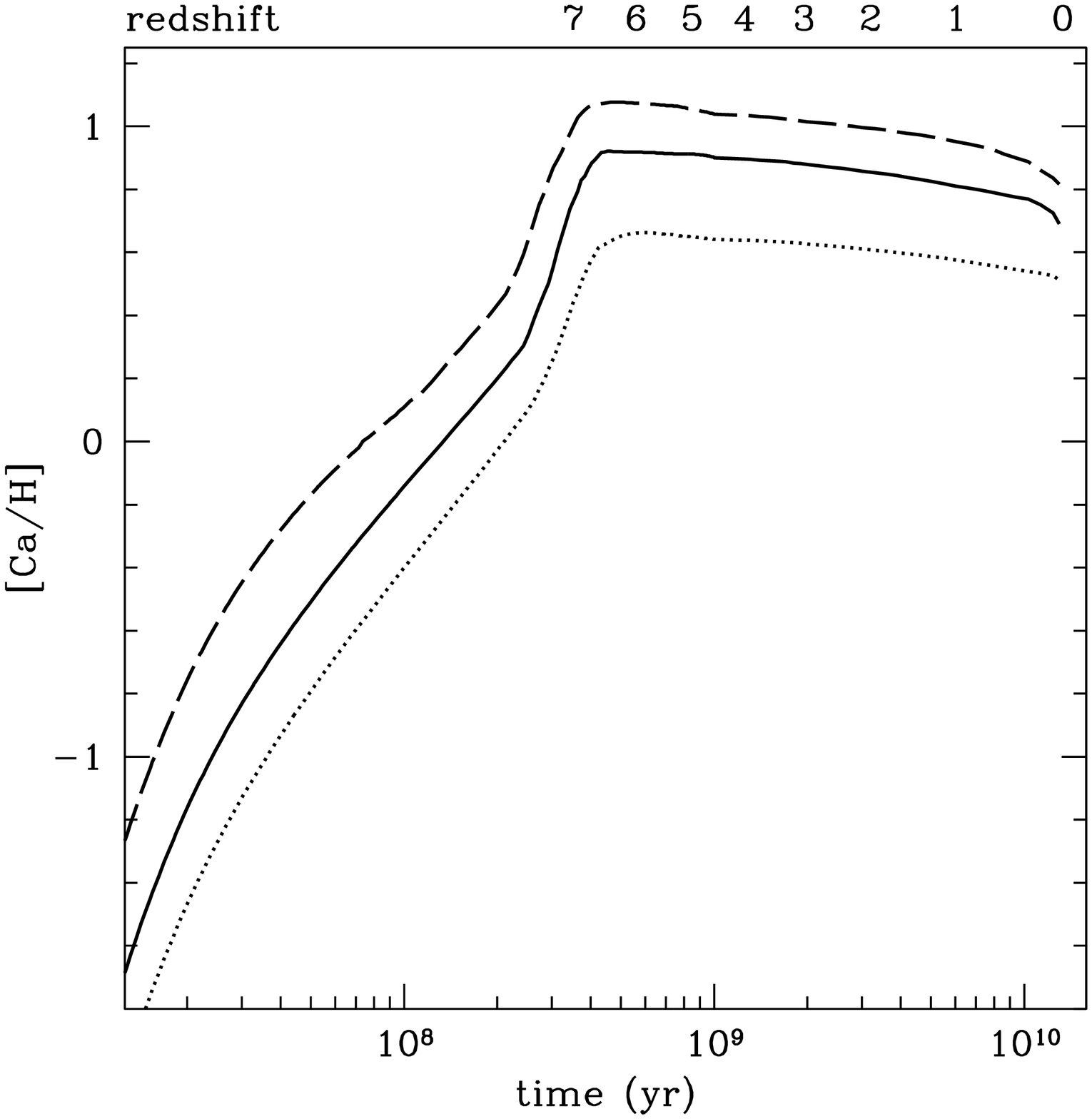}
\includegraphics[width=0.31\textheight]{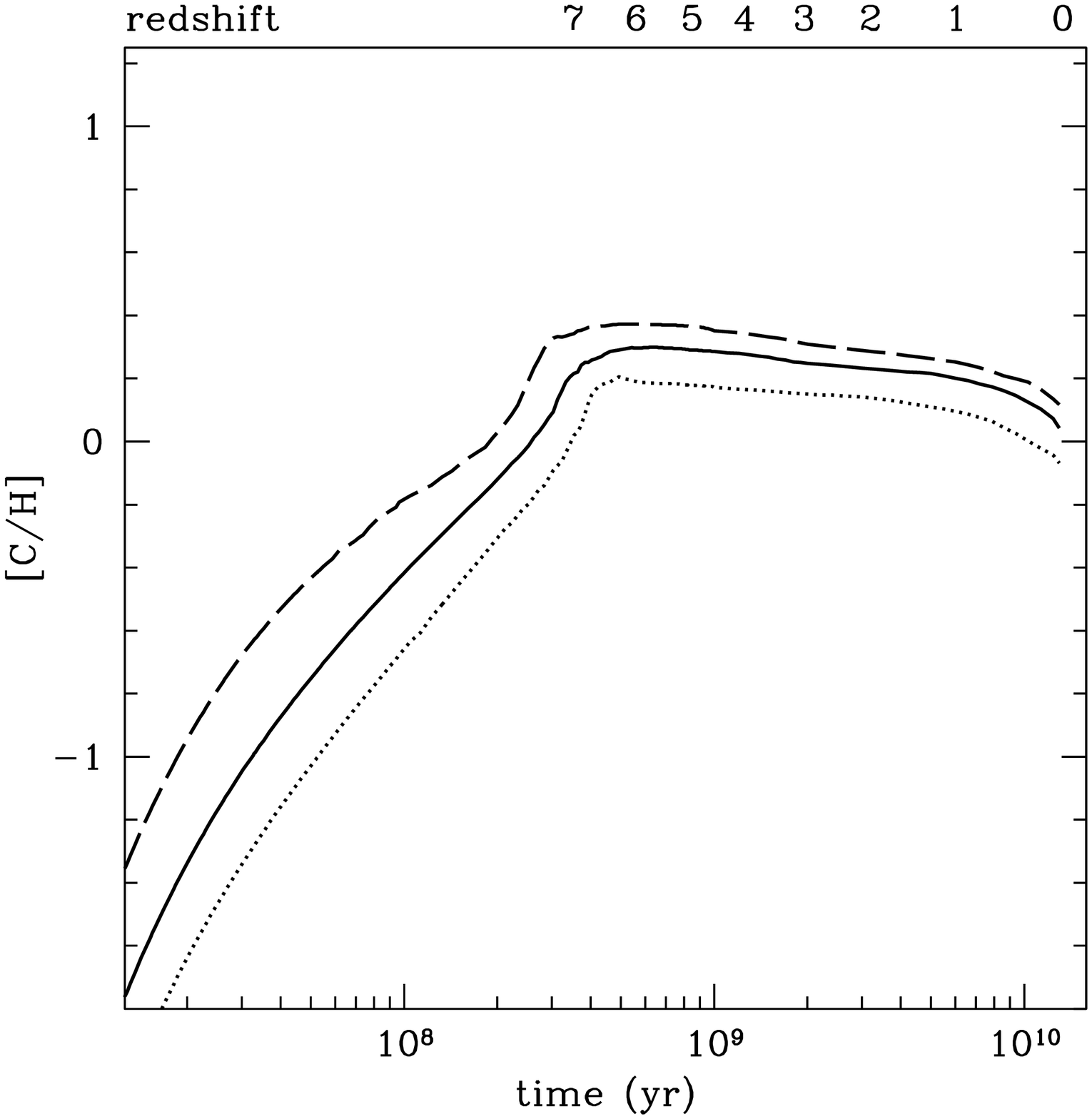}%
\includegraphics[width=0.31\textheight]{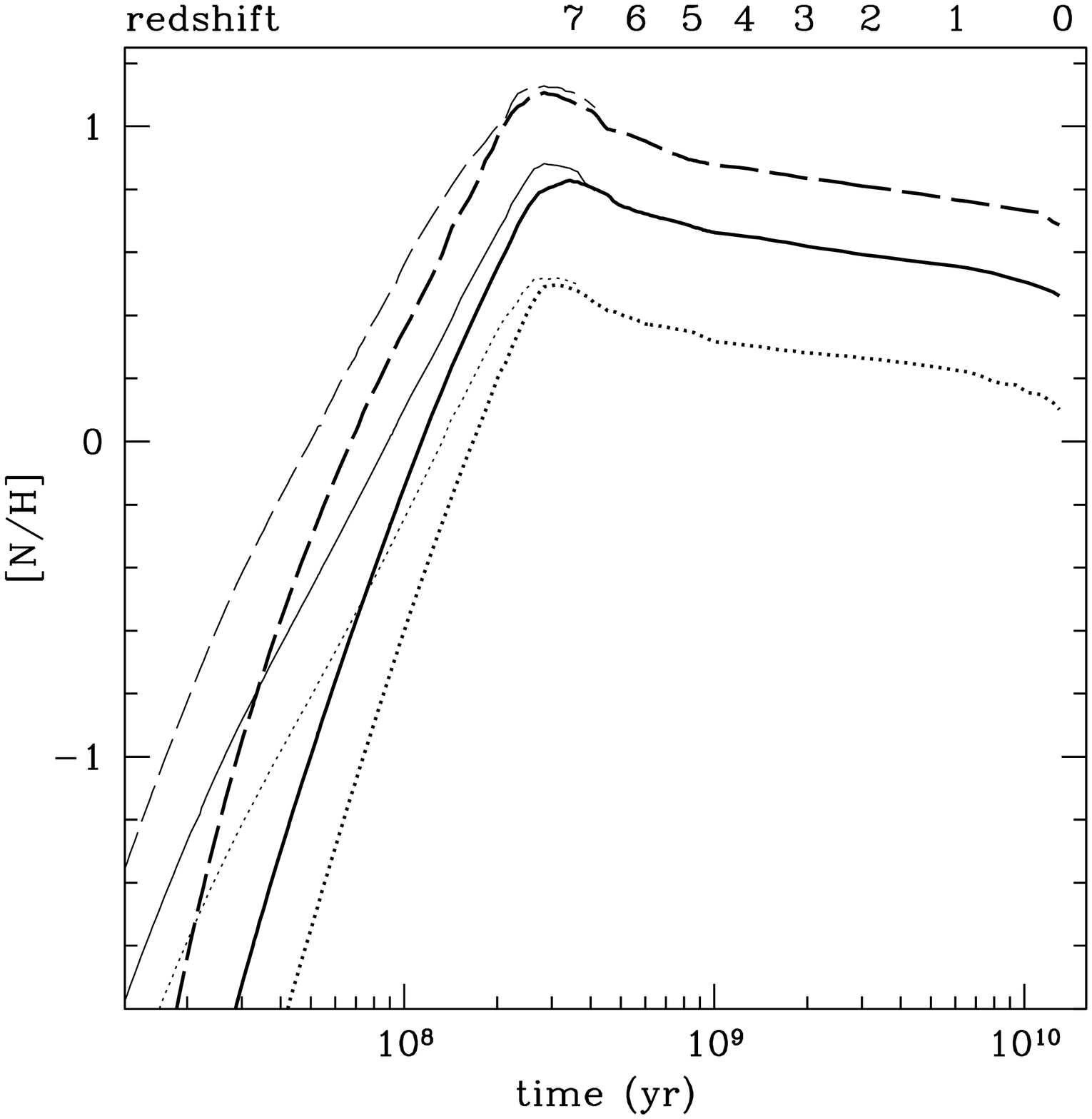}
\caption{Evolution with time and redshift of the [X/H] abundance
  ratios for $\alpha$-elements (O, Mg, Si and Ca), C and N in bulges
  of various masses. 
  The discontinuity at $\sim 2\times 10^{8}$ yr in
  the [Si/H] and [Ca/H] and is due to the occurrence of the galactic
  wind which shortly precedes the maximum in the Type Ia SN rate. 
  The thin lines in the 
  [N/H] \emph{vs.} time plot represent the results obtained by
  adopting primary nitrogen in massive stars as in Matteucci (1986).}
\label{fig:fig09} 
\end{figure*}

The time dependencies of the abundances of the elements under study
for each mass is shown in the Figs. \ref{fig:fig08} (iron) and
\ref{fig:fig09} ($\alpha$-elements, carbon and nitrogen).  A fast
increase in the abundances is noticeable at early times, as well as a
weak decrease at later times, for all elements and all masses; after
the galactic wind, i.e. for $t\gtrsim0.3$ Gyr (which corresponds to a
redshift $z\simeq 6$ in the adopted cosmology) the abundances decrease
by a factor smaller than 2.  Such weak decrease occurs over a period
of more than 13 Gyr.  This can explain the observed constancy of the
QSO abundances as a function of redshift (Osmer \& Shields,~1999).

The elements can be divided according to their behaviour after the wind:
\begin{itemize} 

\item[-] O and Mg show moderate overabundances relative to solar.
They are essentially produced by massive stars ($M > 10M_{\odot}$) on
short timescales ($\lesssim 10^{7}$ years), i.e. almost in lockstep
with the star formation.  After the galactic wind, therefore, their
abundance cannot increase any more because star formation vanishes.
They tend instead to be diluted due to the effect of stellar mass
loss.  In the Galactic bulge, at the time of the onset of the wind, Mg
is more abundant than O, since although [O/H] is higher at earlier
times, it declines very rapidly due to the dependence on metallicity
of the adopted O yields (see Ballero et al., 2007a and references
therein).  Therefore, we expect the mean value attained by [O/H] after
the wind to be smaller than that of [Mg/H], which is what we predict.
In fact, the mean [O/H] for $z \lesssim 6$ ranges from $+0.18$ to
$+0.48$ dex, i.e. oxygen is $\sim1.5$ to 3 times solar, in good
agreement with observations (Storchi-Bergmann et al., 1996; Fraquelli
\& Storchi-Bergmann, 2003); and the mean [Mg/H] ranges from $+0.38$ to
$+0.69$ dex (i.e. 2.5 to 5 times solar), depending on the mass of the
host bulge.  Finally, we notice that both [O/H] and [Mg/H] reach solar
values at times closer to those when $Z=Z_{\odot}$ for the different
models, and therefore are better proxies for the metallicity in the
bulge with respect to [Fe/H].  This is because in the bulge the
metallicity is dominated by O.  

\item[-] Fe, Si and Ca are characterized by a bump immediately after
the occurrence of the galactic wind, which leads to a significant
increase in their abundance, reaching remarkably higher values than in
the previous cases ([Fe/H]: $+0.83$ to $+1.07$, i.e. $\sim$7$-$10
times solar; [Si/H]: $+0.79$ to $+1.13$, i.e. $\sim$6$-$12 times
solar; [Ca/H]: $+0.63$ to $+1.06$, i.e. $\sim$4$-$10 times solar).
This bump is due to the combined effect of the discontinuity caused by
the onset of the galactic wind at times of 0.21 to 0.28 Gyr depending
on the model mass, and of the maximum of the Type Ia SN rate,
occurring at $\sim0.2-0.3$ Gyr.  In fact, these elements are produced
also by Type Ia SNe (Fe mainly, Si and Ca in part, as they are also
produced by Type II SNe) which derive from
progenitors with masses ranging from 0.8 to $8 M_{\odot}$.  These
elements are restored to the ISM on timescales ranging from $\sim30$
Myr to a Hubble time.  So, their abundances could in principle keep
increasing even after the star formation has ceased.  However, due to
the adopted top-heavy IMF ($x_{2}=0.95$), a relatively small fraction
of low- and intermediate-mass stars was produced with respect to what
we would expect if we had adopted a steeper IMF (see below for a
comparison with the IMF of Salpeter, 1955), and therefore what we
observe is a decrease of the Si, Fe and Ca abundances after the
galactic wind.  The estimates for Fe in Seyferts are lower than those
calculated here, but we must take into account several factors which
would lead to an underestimation of the Fe produced by stars
(e.g. depletion by dust; see Calura et al. 2007b).

\item[-] C, as well as O and Mg, is slightly overabundant ($\sim$1.3
to 2.7 times solar). Although it is mostly produced by low- and
intermediate-mass stars, in this same range of masses it is also used
up to form N.  In this case the bump is less pronounced and is mainly
caused by the discontinuity induced by the galactic wind, which
enhances the relative contribution of low- and intermediate-mass
stars.

\item[-] Finally, a behaviour similar to Fe is expected from N, since
the bulk of this element originates from stars with $M<8M_{\odot}$.
There are however two significant differences: first, due its
secondary nature, its abundance increases much more rapidly than other
elements with time (and with metallicity); and second, this very rapid
increase hides the bump at the onset of the wind.  We also adopt a
primary N production in massive stars (Matteucci, 1986; see also
Chiappini et al., 2006).  The effects of this choice are however only
visible before $t_{GW}$, i.e. when nucleosynthesis from massive stars
is active; the decrease of the [N/H] ratio towards earlier times
(i.e. lower metallicities) is less rapid.  Then, after the wind, the
plots with and without primary N essentially overlap.  This was
already observed in Ballero et al. (2006) and Ballero et al. (2007a).
The mean values attained by [N/H] after the galactic wind ranges from
$+0.28$ to $+0.87$, i.e. nitrogen is $\sim$2 to 8 times overabundant
with respect to solar; however, [N/H] reaches overabundances of
$\sim$3$-$10 times solar at its peak. These results are in agreement
with the estimates for Seyferts (Schmitt et al., 1994;
Storchi-Bergmann et al., 1996; Rodr{\'i}guez-Ardila et al., 2005).
\end{itemize}

\begin{figure}
\centering
\includegraphics[width=0.5\textwidth]{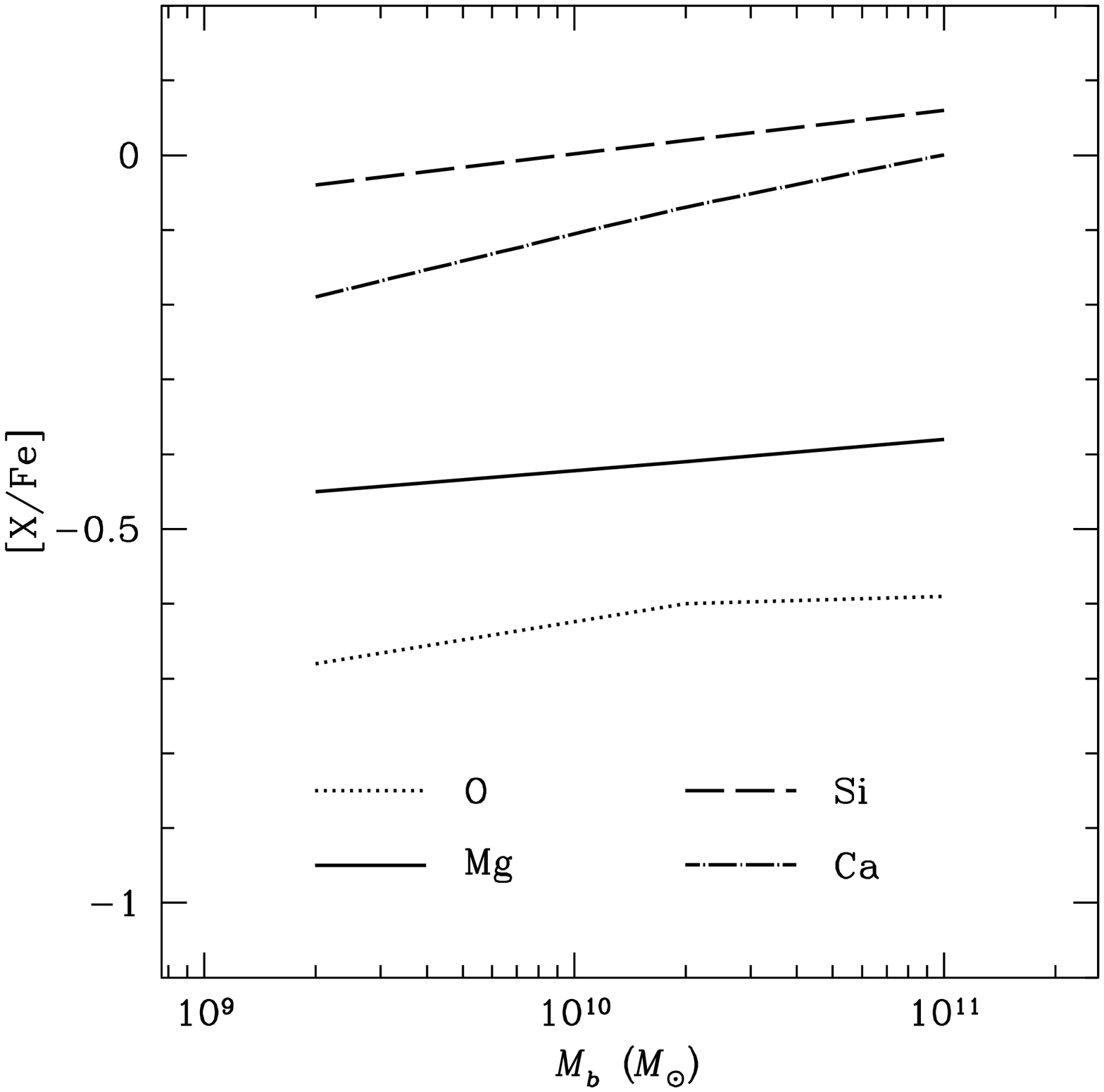}
\caption{Correlation with bulge mass of the [X/Fe] abundance ratios
  for $z \lesssim 6$ for the $\alpha$-elements O, Mg, Si, Ca.} 
\label{fig:fig10} 
\end{figure}

The difference of behaviour among the $\alpha$-elements (O, Mg
\emph{vs.} Si, Ca) is even more evident in Fig. \ref{fig:fig10}, where
the correlation with mass of the [$\alpha$/Fe] abundance ratios after
the wind is shown.  A net separation is present: whereas [Si/Fe] and
[Ca/Fe] are approximately solar, O and Mg are underabundant with
respect to solar, O showing a more pronounced underabundance.  This is
also in agreement with the estimates of a slightly supersolar [Fe/Mg]
in QSOs and of its weak correlation with luminosity (Dietrich et al.,
2003a) Since both [Fe/H] and [Mg/H] are constant within a factor of 2
up to high redshifts ($z \simeq 6$) and follow the same declining
trend with time, we expect this relation to hold for most of the bulge
lifetime, which is what is observed for QSOs (Thompson et al., 1999;
Iwamuro et al., 2002; Freudling et al., 2003; Dietrich et al., 2003a;
Barth et al., 2003; Maiolino et al., 2003; Iwamoto et al., 2004).
Here, we predict [Fe/Mg] values ranging from +0.38 to +0.45, the
highest values corresponding to less massive galaxies where star
formation stops later and Type Ia SNe have more time to pollute the
bulge ISM.

\begin{figure}
\centering
\includegraphics[width=0.5\textwidth]{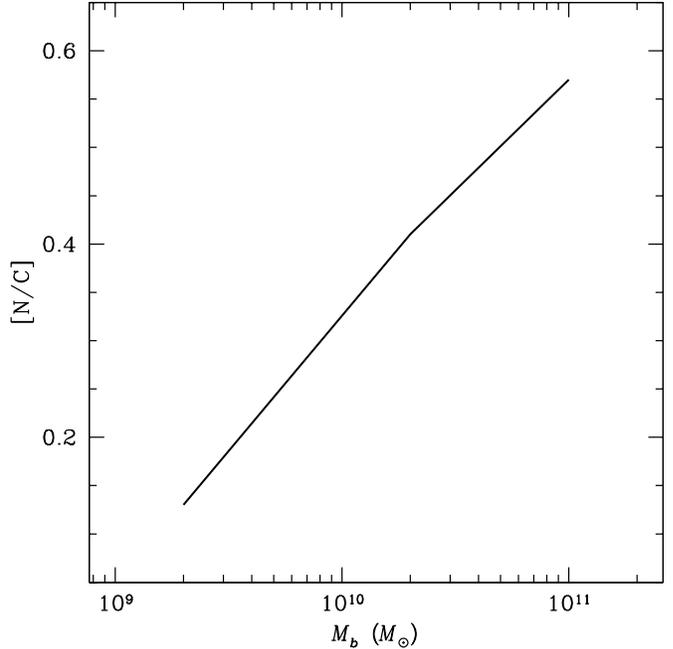}
\caption{Correlation with bulge mass of the [N/C] abundance ratio
  for~$z \lesssim 6$.} 
\label{fig:fig11} 
\end{figure}

Fig. \ref{fig:fig11} shows the variation with mass of the [N/C]
abundance ratio.  The figure illustrates well that this ratio is
remarkably sensitive to the galaxy mass, which, as we have seen in
Fig. \ref{fig:fig07}, is correlated with the galaxy metallicity.  This
happens because of the secondary nature of N, which is produced at the
expense of C in a fashion proportional to the metallicity of the ISM.
The mean [N/C] ratio for $z \lesssim 6$ ranges from +0.13 to +0.57
depending on the bulge mass, i.e. the N/C ratio is $\sim$ 1.4 to 3.7
times solar.  If we consider the amount of variation of the two
abundances, these values are consistent with the estimates of Fields
et al. (2005a).

\begin{figure}
\centering
\includegraphics[width=0.31\textheight]{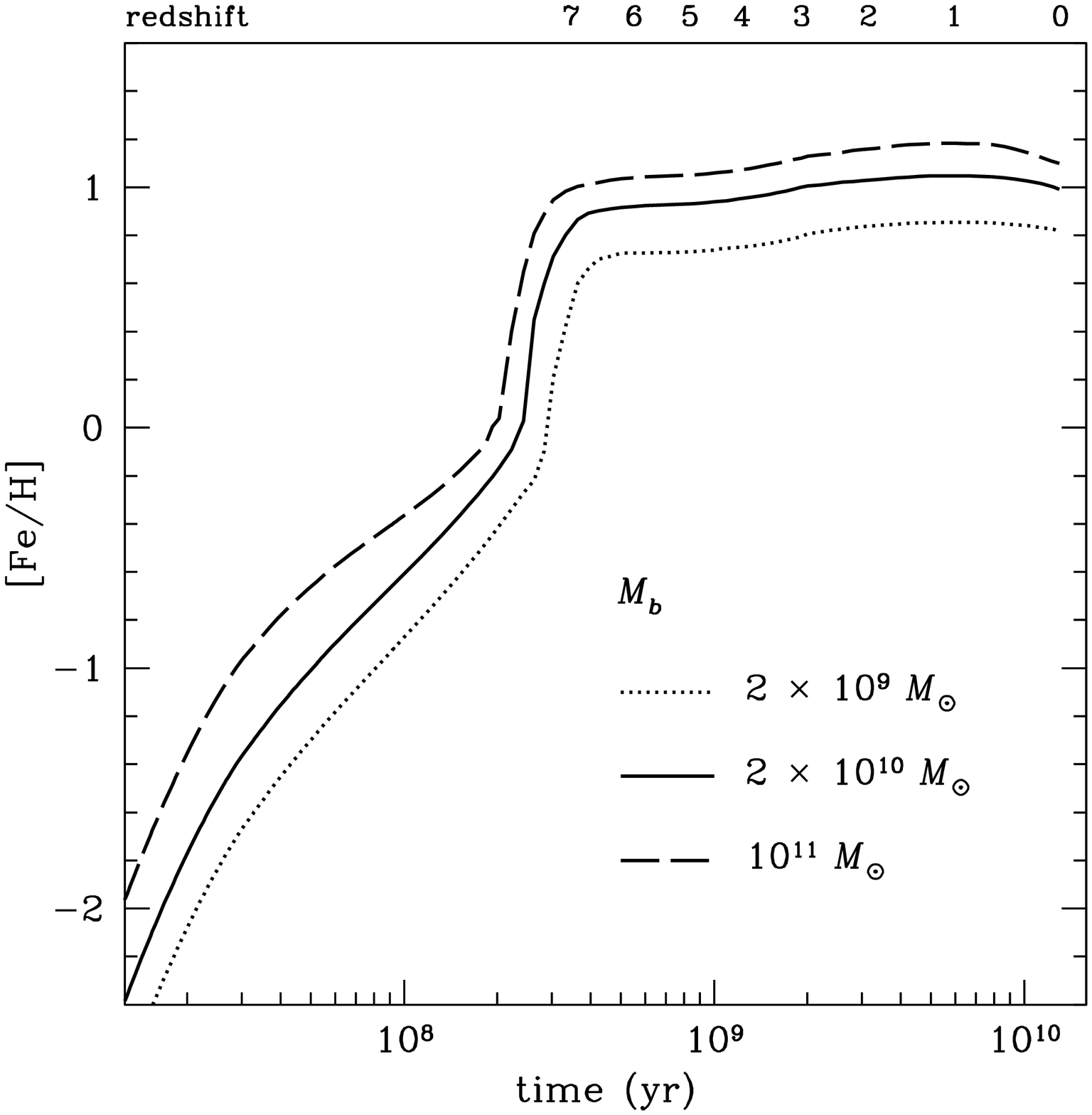}
\includegraphics[width=0.31\textheight]{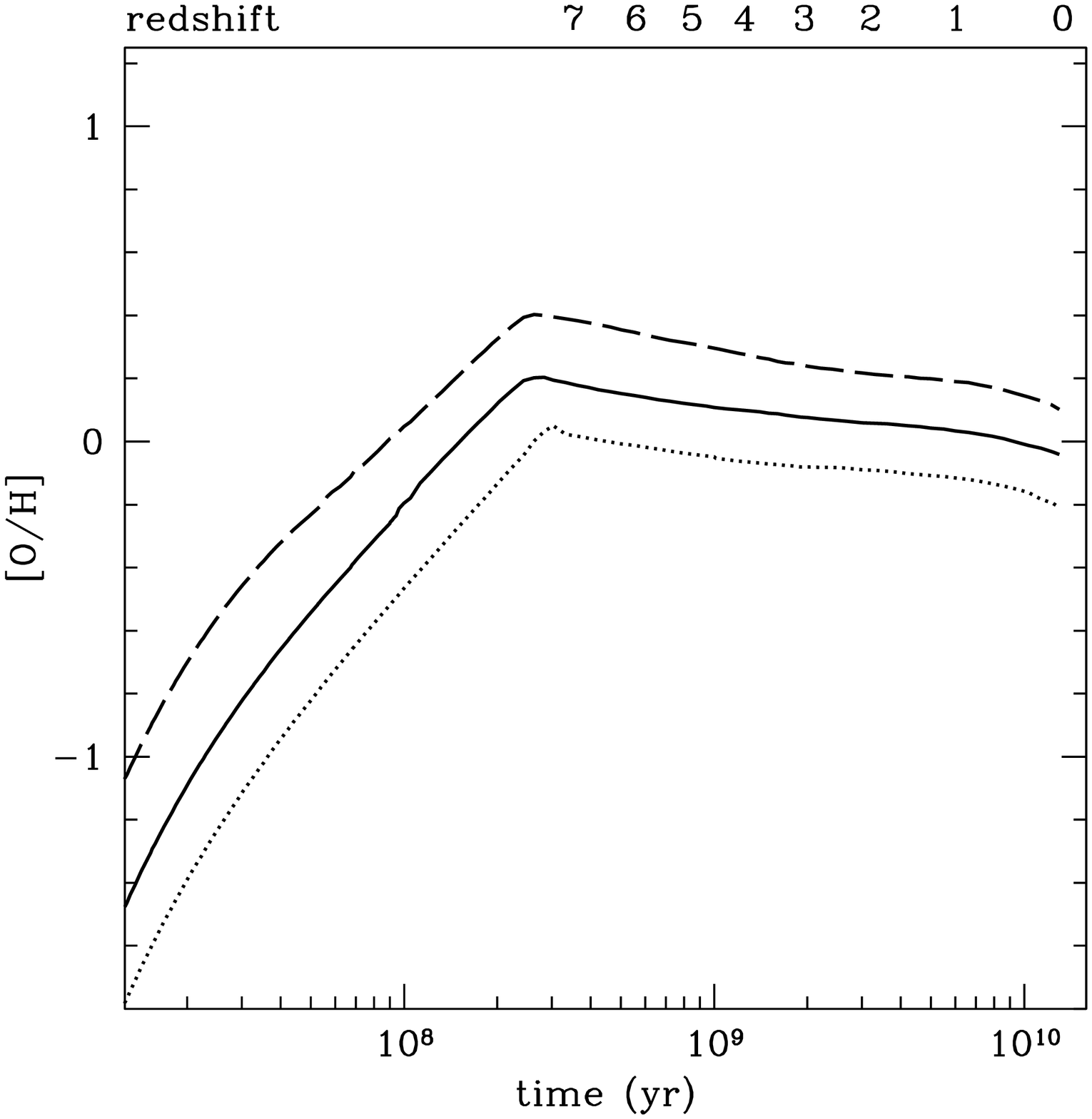}
\includegraphics[width=0.31\textheight]{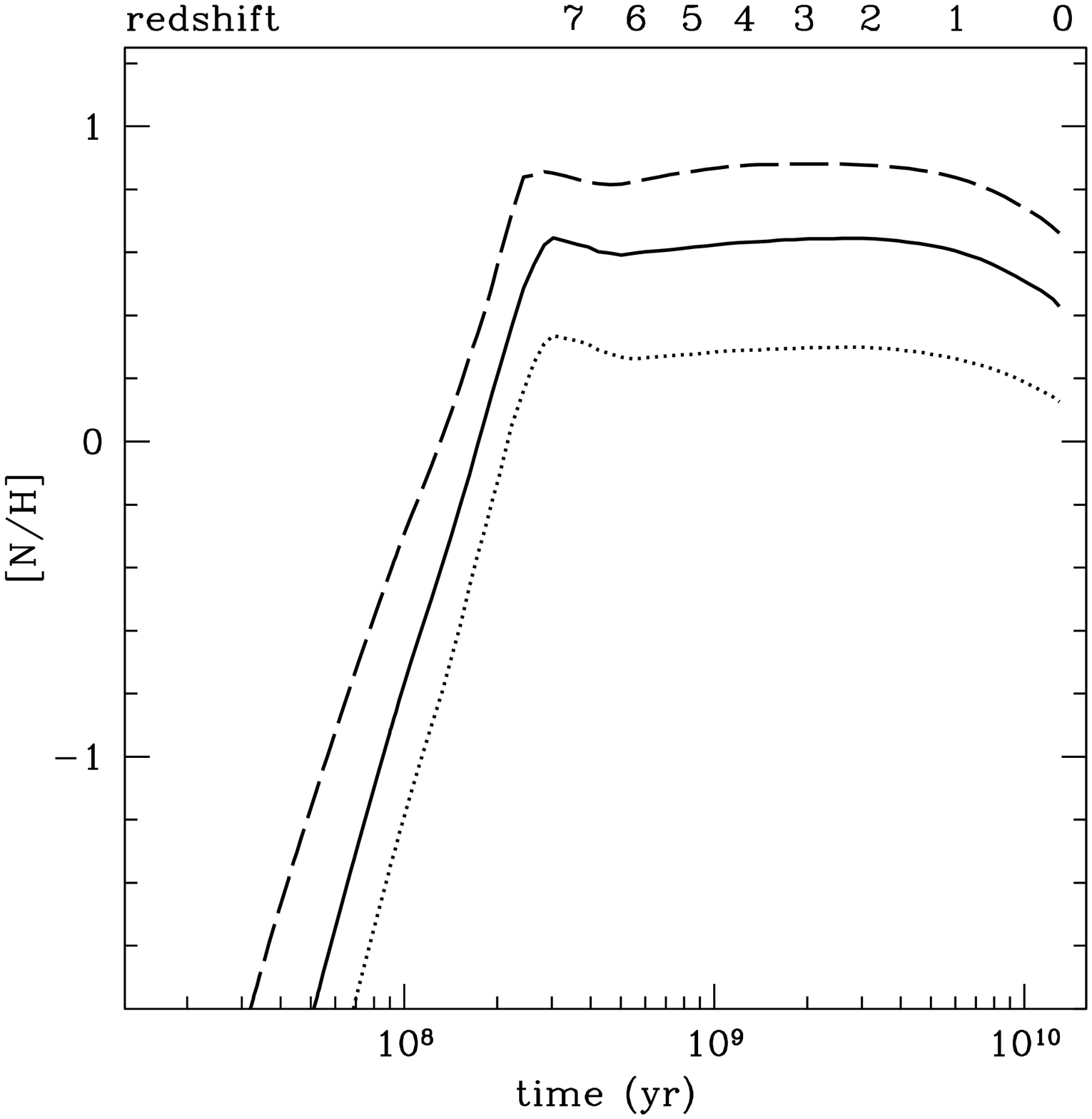}
\caption{Evolution with time and redshift of the [X/H] abundance
  ratios for Fe, Mg and N in bulges of various masses and the adoption
  of a Salpeter (1955) IMF above $1M_{\odot}$. The discontinuity at
  $\sim 3\times 10^{8}$ yr in the [Fe/H] and [Si/H] and is due to the
  occurrence of the galactic wind which shortly precedes the maximum
  in the Type Ia SN rate.} 
\label{fig:fig12} 
\end{figure}

Finally, we make a brief comparison of the results obtained with the
top-heavy IMF ($x_{2}=0.95$) with those we obtain if instead we adopt
$x_{2}=1.35$ (Salpeter, 1955).  In Ballero et al. (2007a) this IMF was
excluded on the basis of the stellar metallicity distribution, since
with our model the Salpeter index for massive stars gives rise to a
distribution which is too metal-poor with respect to the observed ones
(Zoccali et al., 2003; Fulbright et al., 2006), and Ballero et
al. (2007b) further seemed to confirm this point. 
However, Pipino et al. (2007) showed that a hydrodynamical
model for ellipticals can be adapted to the galactic bulge and, with a
spherical mass distribution, it reproduces the above mentioned
metallicity distributions with a Salpeter IMF.  
Furthermore, as we shall show (see \S
\ref{sec:photo}) the $(B-I)$ colors and bulge $K$-band luminosity are
better reproduced by a Salpeter IMF above $1 M_{\odot}$.  Therefore a
brief comparison is useful.

In Fig. \ref{fig:fig12} we show the evolution with time and redshift
of the abundance ratios relative to hydrogen for some of the elements
considered, namely Fe, O and N.  We find similar trends for all the
elements.  However, in the Salpeter case, the abundances at the wind
are lower than with a top-heavy IMF, because the enrichment from
massive stars is lower.  Then, those elements which are produced
mainly by low- and intermediate-mass stars (in this case, Fe and N)
keep increasing after the wind until they reach a maximum value which
is generally slightly higher than that obtained with the top-heavy
IMF, then their abundance stops increasing.  The bump in the [Fe/H] is
$\sim0.1$ dex larger because the production of Type Ia SN progenitors
is favored in this case.  On the contrary, oxygen abundance decreases
steadily after the wind because it is essentially produced by massive
stars and its value remains below the one calculated with the
top-heavy IMF.  The abundance ratios are still almost constant for $z
\lesssim 6$; their mean values are respectively: \begin{itemize} \item
[-] [Fe/H] = 0.79 to 1.08, i.e. 6 to 12 times solar; \item [-] [O/H] =
$-0.09$ to 0.27, i.e. 0.8 to 2 times solar; \item [-] [N/H] = 0.29 to
0.84, i.e. 2 to 7 times solar.  \end{itemize} The values for Fe are
thus again overestimated with respect to observations, but as we said,
the latter are not corrected for Fe depletion by dust or dilution by
dusty torus continuum (Ivanov et al. 2003).  The total metallicity $Z$
ranges from about 3.5 to 6.5 solar, i.e. slightly less than in the
top-heavy IMF case, but still in agreement with observational
estimates, as well as the N and O abundances.

We point out that the adoption of a Salpeter IMF would result into
little change in the quantities calculated previously, i.e. mass
accretion rate, luminosity, energetics and final black hole mass.  The
accretion rates of Eddington and Bondi do not depend on the adopted
IMF.  A steeper IMF only slightly shifts the time of occurrence of the
wind of a few tens Myrs ahead, thus prolonging the Eddington-limited
accretion phase which, as we said, is the phase when most of accretion
and shining occurs.  This will lead to a larger BH mass; however, the
final BH masses increase by only about $10\%$.

\subsection{Bulge colours and the colour-magnitude relation}
\label{sec:photo}

In this section, we present our results for the spectro-photometric
evolution of the three bulge models studied in this paper.
The photometrical evolution of Seyfert galaxies, which requires
modeling of the AGN continuum and, for Type 2 Seyferts, of the dusty
torus, is not treated at present and may be the subject of a
forthcoming paper.

\begin{figure}
\includegraphics[width=.5\textwidth]{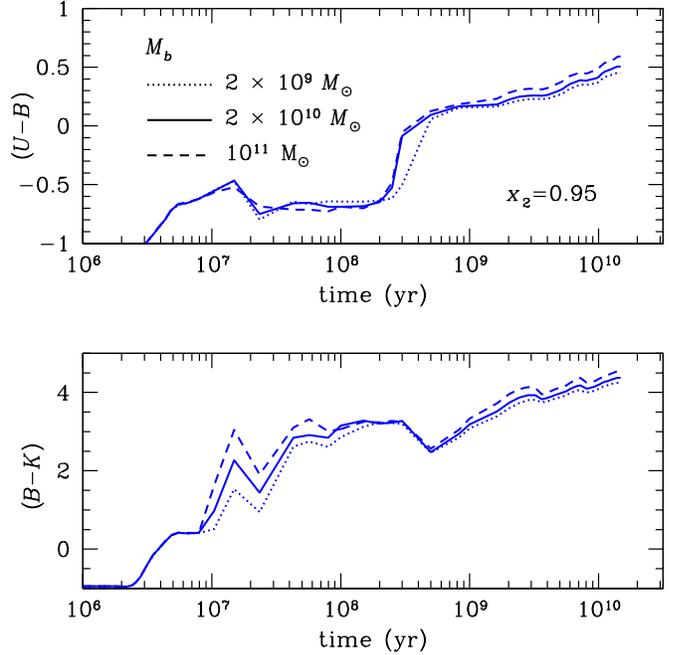}
\caption{Evolution of the predicted $(U-B)$ (upper panel) and $(B-K)$
  (lower panel) colours for bulge models of three different masses
  (\emph{dotted line:} $M_{b}=2\times 10^{9} M_{\odot}$; \emph{solid
    line:} $M_{b}=2 \times 10^{10} M_{\odot}$; \emph{dashed line:}
  $M_{b}=10^{11} M_{\odot}$), assuming an IMF with  $x_{2} = 0.95$.} 
\label{fig:fig13} 
\end{figure}

In Fig. \ref{fig:fig13}, we show the predicted time evolution of the
\mbox{$(U-B)$} and $(B-K)$ colours for the three bulge models studied
in this work.  In Figure 1, we assume an IMF with $x_{1} = 0.33$ for
stars with masses $M$ in the range $0.1\le M/M_{\odot} \le 1$ and
$x_{2} = 0.95$ for $M>1 M_{\odot} $.  At all times, higher bulge
masses correspond to redder colours, owing both to a higher
metallicity and to an older age.  This is consistent with the popular
``downsizing'' picture of galaxy evolution, according to which the
most massive galaxies have evolved faster than the less massive ones
(Matteucci 1994, Calura et al. 2007c).

\begin{figure}
\includegraphics[width=.5\textwidth]{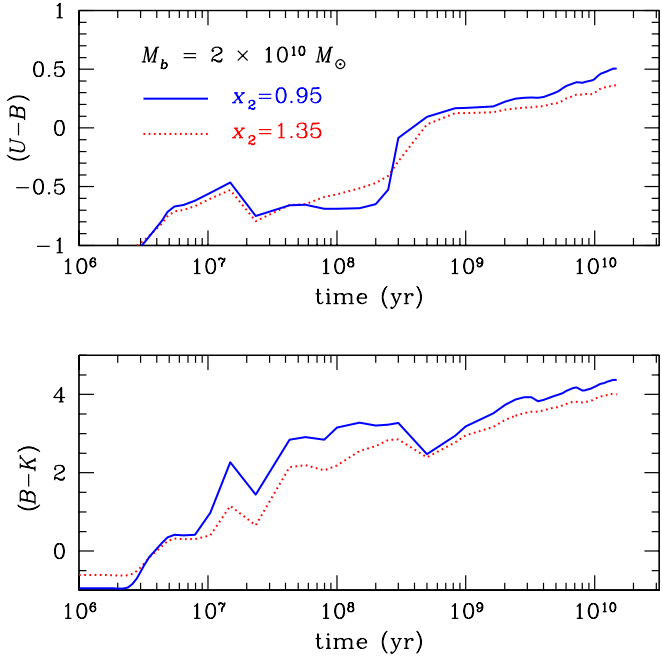}
\caption{Evolution of the predicted $(U-B)$ (upper panel) and $(B-K)$
  (lower panel) colours for a bulge model of mass $M_{b}=2 \times
  10^{10} M_{\odot}$ assuming two different IMFs.  
The solid lines represent the colours computed by assuming $x_2=0.95$,
whereas the dotted lines are computed assuming  $x_2=1.35$.}  
\label{fig:fig14}
\end{figure}

In Fig. \ref{fig:fig14}, we show how the assumption of two different
IMFs affects the predicted evolution of the $(U-B)$ and $(B-K)$
colours, for a bulge of mass $M_{b} = 2 \times 10^{10}
M_{\odot}$. We compare the colours calculated with the IMF presented
in \S \ref{sec:chem} with the ones calculated by assuming an IMF with
$x_{1} = 0.33$ for stars with masses $M$ in the range $0.1\le
M/M_{\odot} \le 1$ and $x_{2} = 1.35$ for $M>1 M_{\odot} $.  It is
interesting to note how, at most of the times, the assumption of a
flatter IMF for stars with masses $m>1 M_{\odot} $ implies redder
colours. This is due primarily to a metallicity effect, since a
flatter IMF implies a larger number of massive stars and a larger
fraction of O (the element dominating the metal content) restored into
the ISM at all times.

\begin{figure}
\includegraphics[width=.5\textwidth]{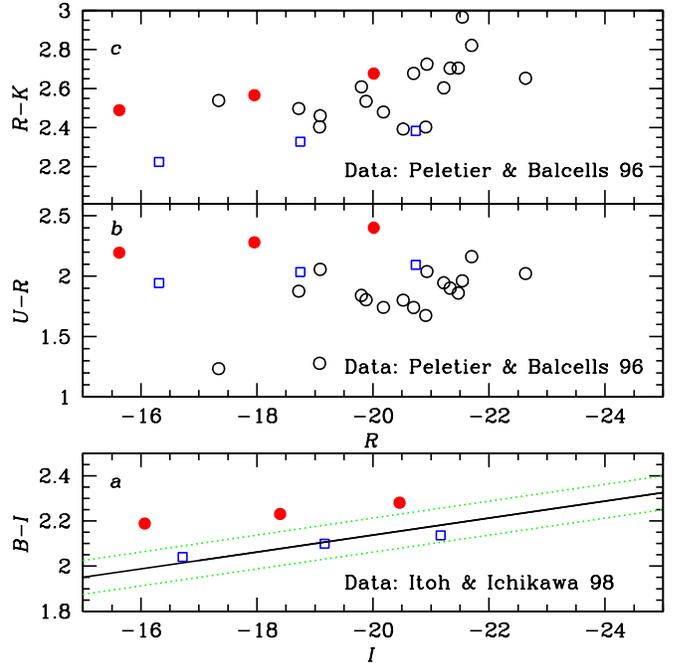}
\caption{Predicted and observed colour-magnitude relation for bulges. 
The solid line and the dotted lines are the regression and the
dispersion of CMR observed by Itoh \& Ichikawa (1998), respectively.  
The solid circles and open squares are our predictions for the three
bulge models studied in this paper, computed by assuming for the  
IMF $x_2=0.95$ and $x_2=1.35$, respectively. 
Panel $b$: $R$ \emph{vs.} $(U-R)$ diagram. 
The open circles are the data by Peletier \& Balcells (1996). 
Solid circles and open squares as in panel $a$. 
Panel $c$: $R$ \emph{vs.} $(R-K)$ diagram. 
Open circles, solid circles and open squares as in panel $a$.} 
\label{fig:fig15} 
\end{figure}

In Figure \ref{fig:fig15}, we show the colour magnitude relation (CMR)
predicted for our bulge models and compared to the data from Itoh \&
Ichikawa (1998, panel $a$), and Peletier \& Balcells (1996, panels $b$
and $c$).

Itoh \& Ichikawa measured the colours of 9 bulges in a fan-shaped 
aperture opened along the minor axis, in order to minimize the effects
of dust extinction.
In panel $a$ of Fig. \ref{fig:fig15} we show the linear regression to
the data observed by Itoh \& Ichikawa (1998), indicated by the solid
line, whereas the two dotted lines express the dispersion.  

Peletier \& Balcells (1996) determined the colours for a sample of
local bulges, for which they estimated that the effect of dust
reddening is negligible.
Of the sample studied by Peletier \& Balcells (1996), here we consider
a subsample of 17 bulges, for which Balcells \& Peletier (1994)  have
published the absolute $R$ magnitudes. This allows  
us to plot an observational CMR.

The predicted CMR has been calculated by
adopting two different IMFs, i.e. with $x_{2}=0.95$ and $x_{2}=1.35$,
represented in Fig. \ref{fig:fig15} by the solid circles and the open
squares, respectively.  
From the analysis of the $I$ \emph{vs.} $(B-I)$ plot, we note that the
adoption of a flatter IMF leads to an 
overestimation of the predicted $(B-I)$ colours; in fact, the
predictions for the flatter IMF lie above the upper dotted lines,
representing the upper limits for the data observed by Itoh \&
Ichikawa (1998).  
On the other hand, the predictions computed with an
IMF with $x_2=1.35$ are consistent with the available observations.

From the analysis of the $R$ \emph{vs.} $(U-R)$ diagram, 
we note that our predictions computed assuming  $x_{2}=1.35$ are
consistend with the observations, in particular for the bulge models
of masses  $2\times 10^{10} M_{\odot}$ and $10^{11} M_{\odot}$.
For the lowest mass bulge, corresponding to the highest absolute R
magnitude, the $(U-R)$ colour seems to be overestimated. 
In Table \ref{tab:tab3}, we present our results for the predicted
present day colours for the bulge model of mass $2 \times 10^{10}
M_{\odot}$, computed for two different IMFs, compared to observational
values for local bulges derived by various authors.  The present-day
values are computed at 13 Gyr.  Note that, for each colour, the
observational values represent the lowest and highest observed values
reported by the authors.  The values we predict are compatible with
the observations of local bulges, which show a considerable spread.
However, 
the sample of bulges considered here seems to be skewed towards high
mass bulges, hence it is difficult to infer the trend of the  $R$
\emph{vs.} $(U-R)$ for magnitudes $R<-16$ mag.
On the other hand, the predictions computed by assuming $x_{2}=0.95$
seem to produce too much high $(U-R)$ colours with respect to the
observations. 

Finally, the $R$ \emph{vs.} $(R-K)$ diagram does not allow us to draw
any firm conclusion on the slope of the IMF.

From the combined study of the $I$ \emph{vs.} $(B-I)$, $R$ \emph{vs.}
$(U-R)$  and $R$ \emph{vs.} $(R-K)$ diagrams, we conclude that the
observational data for local bulges  seem to disfavour IMF flatter  
than $x_{2}=1.35$ in the stellar mass range $1 \leq M/M_{\odot} \leq 80$. 
In Table \ref{tab:tab3}, we present our results for the predicted
present day colours for the three bulge models presented in this paper,  
computed for two different IMFs. 
The present-day values are computed at 13.7 Gyr.  
In the same table, we show the observational 
data of Peletier \& Balcells (1996) and with the one of Galaz et
al. (2006), who estimated that for their sample, dust effects should
be small, with an upper limit on the extinction of 0.3 mag.  
Note that, for each colour, the
observational values represent the lowest and highest observed values
reported by the authors. 

\begin{table}
\centering
{\bf Photometric properties of the Galactic Bulge}\\
\ \\
\footnotesize
\begin{tabular}{cl|l|l|l}
\hline
\hline
&  & $(B-R)$ & $(U-R)$ & $(R-K)$\\
\hline
Observed &($a$)     & $1.27 - 1.6$  & $1.24 - 2.16$ & $2.39 - 2.97$ \\
         &($b$)     & $0.5 - 2.18$  &               &               \\
\hline
Predicted&($x=1.35$)& $1.64 - 1.71$ & $1.94 - 2.09$ & $2.22 - 2.38$ \\
         &($x=0.95$)& $1.75 - 1.83$ & $2.20 - 2.40$ & $2.49 - 2.68$ \\
%
\hline
\hline
\noalign{\smallskip}
\end{tabular}
\caption{Predicted colours for the Galactic Bulge assuming two different IMFs,
compared to observational values of local bulges from various sources
($a$: Balcells \& Peletier 1994; $b$: Galaz et al. 2006.)}
\label{tab:tab3}
\end{table}

For the model with mass comparable to the one of the Bulge of the
Milky Way galaxy, i.e. the one with a total mass \mbox{$M = 2 \times
10^{10} M_{\odot}$}, with the assumption of an IMF with $x_{2}=0.95$ we
predict a present $K$-band luminosity $L_{Bul,K} = 4.5 \times 10^{9}
L_{\odot}$.  Existing observational estimates of the $K$-band
luminosity of the bulge indicate values $0.96 \times 10^{10} \le
L_{Bul,K,obs} \le 1.8 \times 10^{10} L_{\odot}$ (Dwek et al. 1995,
Launhardt et al. 2002), i.e. al least a factor of $2$ higher than the
estimate obtained with our model.  By adopting an IMF with
$x_{2}=1.35$, we obtain $L^{x=1.35}_{Bul,K} = 7.6 \times 10^{9}
L_{\odot}$, in better agreement with the observed range of values.

\section{Summary and conclusions}

We made use of a self-consistent model of galactic evolution which
reproduces the main observational features of the Galactic bulge
(Ballero et al., 2007a) to study the fueling and the luminous output
of the central supermassive BH in spiral bulges, as fed by the stellar
mass loss and cosmological infall, at a rate given by the minimum
between the Eddington and Bondi accretion rates.  A realistic
galaxy model was adopted to estimate the gas binding energy in the
bulge, and the combined effect of AGN and supernova feedback was taken
into account as contributing to the thermal energy of the interstellar
medium.  We also investigated the chemical composition of the gas
restored by stars to the interstellar medium.  Assuming that the gas
emitting the broad and narrow lines observed in Seyfert spectra is
well mixed with the bulge interstellar medium, we have made specific
predictions regarding Seyfert metallicities and abundances of several
chemical species (Fe, O, Mg, Si, Ca, C and N), and their redshift
evolution.  Finally, we calculated the evolution of the $(B-K)$ and
$(U-B)$ colours, the present-day bulge colours and $K$-band luminosity
and the color-magnitude relation, and discussed their dependence on
the adopted IMF.

Our main results can be summarized as follows:

\begin{enumerate}

\item The AGN goes through a first phase of Eddington-limited
accretion and a second phase of Bondi-limited accretion.  Most of the
fueling (and of the shining) of the AGN occurs at the transition
between the two phases, which coincides approximately with the
occurrence of the wind.  Within a factor of two, the final black hole
mass is reached in a fraction of the bulge lifetime ranging from 2 to
5\%.

\item The peak bolometric luminosities predicted for AGNs
residing in the bulge of spiral galaxies are in good agreement
with those observed in Seyfert galaxies ($\simeq 10^{42}-10^{44}$ erg
s$^{-1}$).  
At late times, the model nuclear luminosities
(produced by accretion of the mass return from the passively aging
stellar populations) are $<10^{42}$ erg s$^{-1}$, and would be further
reduced by a factor $\sim 0.01$ when considering ADAF (or its
variants) accretion regimes.
To recover an agreement with the luminosities of local Seyferts, it is
necessary to assume a rejuvenation event in the past $1-2$ Gyrs.

\item The feedback from the central AGN is not in any case the main
responsible for triggering the galactic wind, since its contribution
to the thermal energy is at most comparable to that of the supernova
feedback.

\item The proportionality between the mass of the host bulge and that
of the central black hole is very well reproduced without the need to
switch off the accretion \emph{ad hoc}.  We derive an approximate
relation of $M_{BH} \approx 0.0009 M_{b}$, consistent with recent
estimates (McLure \& Dunlop 2002; Marconi \& Hunt 2003; H\"aring \&
Rix 2004).

\item The massive amounts of star formation (several thousands solar
masses a year) necessary to produce the observed line strengths, dust
content and metallicities of Seyfert galaxies are easily attained.
Due to such high star formation rate, solar metallicity is reached in
less than $10^8$ yr, and solar abundances for the elements we
considered are reached in a few hundred million years.  This naturally
explains the high metallicities and abundances inferred for Seyfert
galaxies in analogy with QSOs.

\item After the first $\sim$300 million years, the interstellar medium
in the bulge reaches overabundances of up to one order of magnitude
for N, Fe, Si, Ca, up to 5 times solar for N and 3 times solar for C
and O.  This is roughly consistent with the estimates for Fe, N and O
in the broad and intrinsic narrow line regions of Seyfert galaxies.
The slightly supersolar [Fe/Mg] and its weak dependence on the galaxy
luminosity (mass) are recovered.  These results remain essentially
unchanged if we adopt a Salpeter (1955) stellar initial mass function.

\item The mean value assumed by the [N/C] abundance ratio is very
sensitive on the galaxy mass, due to the sensitivity of N to
metallicity.  The calculated values are consistent with recent
estimates (Fields et al., 2005a).

\item Higher bulge masses imply redder colours, in agreement with the
downsizing picture of galaxy evolution. A flatter IMF gives rise to a
redder bulge. While the present-day bulge colours are well fitted both
by a top-heavy ($x_{2}=0.95$) and a Salpeter ($x_{2}=1.35$) IMF, the latter is
required in order to achieve a better agreement with the
colour-magnitude relation and the present-day $K$-band luminosity of
the bulge.

\end{enumerate}

\acknowledgements
SKB wishes to acknowledge Antonio Pipino for interesting suggestions 
and discussions.
\normalsize

\appendix

\section{The disk gravitational energy contributions}
\label{sec:A1}

In this paper, in order to estimate the binding energy of the gas in
the bulge, we assumed spherical symmetry for the gas distribution
itself. Thus, the contribute of the spherical stellar bulge and the
spherical dark matter halo are given by the elementary expressions
(17) and (18).

A more complicate case is represented by the disk-gas interaction. In
fact, one could suppose that a 2-dimensional integral, involving
special functions must be evaluated, due to the disk
geometry. However, this is not the case: in fact, if each gas element
is displaced radially from $r$ to $\rt$, then
\begin{eqnarray}
\Delta\Egd & = &
\int_0^{\rt}\rho_g(r)r^2dr 
\int_{4\pi}d\Omega[\phi_d(\erv \rt)-\phi_d(\erv r)] =\cr
&=& 4\pi \int_0^{\rt}\rho_g(r)r^2 [\bar{\phi}_d(\rt)-\bar{\phi}_d(r)]dr
\end{eqnarray}
where
\begin{equation}
\bar{\phi}_{d}(r) = {1\over 4\pi}\int_{4\pi}\phi_d(\erv r)d\Omega
\end{equation}
is the angular averaged potential, and $\erv$ is the radial unit
vector.  It is trivial to prove by direct evaluation that the angular
mean of the potential of a generic density distribution is also the
(spherical) potential of the angular averaged density. Of course, this
result can also be obtained in a more elegant way by considering the
angular mean of the expansion of the potential in spherical harmonics
(e.g., Binney \& Tremaine 1987), so that all the terms, except the
monopole, vanish. An identical argument holds also in the case of the
computation of the virial trace. In fact, from $\xv=\erv r$ it follows
that
\begin{equation}
\Wbd = -\int\rho_g <\xv,\nabla\phi_d> d^3\xv
=-4\pi\int\rho_g(r){d\bar\phi_d (r)\over dr}r^3 dr
\end{equation}
The surface density of a razor-thin axisymmetric disk in spherical 
coordinates is
\begin{equation}
\Sigma(r,\vartheta,\varphi) =
\frac{\Sigma_d(r\sin\vartheta)\delta(\cos\vartheta)}{r}
\end{equation}
so that
\begin{equation}
\bar{\rho}_d(r) = {1\over 4\pi}\int_{4\pi}\Sigma(r,\vartheta,\varphi)d\Omega =
{\Sigma_d(r)\over 2r}.
\end{equation}
For the exponential disk in equation (14) we then obtain 
\begin{equation}
\bar{\phi}_d(r) = -{GM_d\over r}\left(1-e^{-r/\Rd}\right),
\end{equation}
so that integrations in (A1) and (A3) reduce to
\begin{eqnarray}
\Delta\widetilde{\Egd} &=& \frac{\delta +(1-h)(1- e^{-\delta/h})}{1+\delta} +\cr
&+&{e^{1/h}\over h}\left[{\rm E_i}\left(-{1\over h}\right)-
{\rm E_i}\left(-{1\over h}-{1\over\delta}\right)\right]
\end{eqnarray}
and
\begin{equation}
\widetilde{\Wbd} = -{1\over h} -{e^{1/h}(1+h)\over h^2}
         {\rm E_i}\left(-{1\over h}\right)
\end{equation}
where $h\equiv\Rd/\rb$, $\delta\equiv\rt/\rb$, and ${\rm
E_i}(-x)=-\int_r^{\infty}e^{-t}t^{-1}dt$ is the Exponential Integral.

\end{document}